\documentclass{article}

\PassOptionsToPackage{numbers,sort&compress}{natbib}
\usepackage{booktabs} % For formal tables
\usepackage[noend,ruled,noline]{algorithm2e} % For algorithms

\usepackage{enumitem}
\usepackage{mathtools}
\usepackage{amsmath,amsthm}
\usepackage{tikz}
\usetikzlibrary{shapes.geometric}

\SetAlFnt{\small}
\SetAlCapFnt{\small}
\SetAlCapNameFnt{\small}
\SetAlCapHSkip{0pt}
\IncMargin{-\parindent}

\usepackage{xcolor}
\usepackage{color-edits}
\addauthor{eb}{blue}
\addauthor{vg}{red}
\usepackage{soul}
\definecolor{DarkOliveGreen}{HTML}{556B2F}
\definecolor{Chocolate}{HTML}{D2691E}
\definecolor{RoyalBlue}{HTML}{4169E1}
\usepackage[main, final]{neurips_2025}
\renewcommand\citep[1]{\citealp{#1}}
\renewcommand\citet[1]{\citeauthor{#1} [\citealp{#1}]}

\DeclareMathOperator*{\argmin}{arg\,min}

\newtheorem{theorem}{Theorem}[section]

\newtheorem{lemma}[theorem]{Lemma}
\newtheorem{observation}[theorem]{Observation}

\newtheorem*{definition*}{Definition}

\newtheorem{corollary}[theorem]{Corollary}

\theoremstyle{remark}

\usepackage{thm-restate}

\usepackage[utf8]{inputenc} % allow utf-8 input
\usepackage[T1]{fontenc}    % use 8-bit T1 fonts
\usepackage[hidelinks]{hyperref}       % hyperlinks
\usepackage{url}            % simple URL typesetting
\usepackage{booktabs}       % professional-quality tables
\usepackage{amsfonts}       % blackboard math symbols
\usepackage{nicefrac}       % compact symbols for 1/2, etc.
\usepackage{microtype}      % microtypography
\usepackage{xcolor}         % colors

\title{Procurement Auctions with Predictions:\\ Improved Frugality for Facility Location}

\author{%
    \begin{tabular}{c@{\hspace{1.5cm}}c}    
        \begin{tabular}{c}
            Eric Balkanski\\
            Columbia University\\
            \texttt{eb3224@columbia.edu}
        \end{tabular}
        &    
        \begin{tabular}{c}
            Nicholas DeFilippis\\
            New York University\\
            \texttt{nad9961@nyu.edu}
        \end{tabular}
        \\
        \\
        \begin{tabular}{c}
            Vasilis Gkatzelis\\
            Drexel University,\\
            Archimedes, Athena R.C., Greece\\
            \texttt{gkatz@drexel.edu}
        \end{tabular}
          & 
          \begin{tabular}{c}
            Xizhi Tan\thanks{This work was done while the author was a PhD student at Drexel University}\\
            Stanford University\\
            \texttt{xizhi@stanford.edu}
        \end{tabular}
    \end{tabular}
}

\begin{document}
\newcommand{\predict}{\hat{o}}
\newcommand{\err}{\eta}
\newcommand{\approxratio}{r}

\def \strp {\mathbf{s}}
\def \str {s}
\def \t {\mathbf{t}}
\def \p {\mathbf{p}}
\def \F {\mathcal{F}}
\def \H {\mathcal{H}}
\def \R {\mathbb{R}}
\def \N {\mathbb{N}}

\newcommand{\robust}{\beta}
\newcommand{\consist}{\alpha}

\newcommand{\minmaxsingle}{\textsc{MinMaxP }}
\newcommand{\minmaxmech}{\textsc{Minimum Bounding Box }}
\newcommand{\minsummech}{\textsc{Coordinatewise Median with Predictions }}

\newcommand{\vecb}{\mathbf{b}}
\newcommand{\vect}{\mathbf{t}}
\newcommand{\vecp}{\mathbf{p}}
\newcommand{\vecx}{\mathbf{x}}

\newcommand{\calN}{\mathcal{N}}
\newcommand{\calB}{\mathcal{B}}
\newcommand{\calT}{\mathcal{T}}
\newcommand{\calP}{\mathcal{P}}

\newcommand{\pgreedym}{\hat{\imath}^{g}}
\newcommand{\poptm}{\hat{\imath}}
\newcommand{\greedym}{i^{g}}
\newcommand{\optm}{i^\star}

\newcommand{\pmake}{\hat{\texttt{tOPT}}}
\newcommand{\mech}{\mathcal{M}}
\newcommand{\simplemech}{\textsc{SimpleScaledGreedy}}
\newcommand{\errbound}{\bar{\eta}}
\newcommand{\algo}{\textsc{Makespan-Min}}
\newcommand{\ALG}{\texttt{ALG}}
\newcommand{\OPT}{\texttt{OPT}}
\newcommand{\pred}{\hat{\mathbf{\open}}}
\newcommand{\tOPT}{\texttt{OPT}}
\newcommand{\ms}{\texttt{MS}}

\newcommand{\instance}{\mathcal{I}}
\newcommand{\allo}{\mathbf{x}}
\newcommand{\pay}{\mathbf{p}}
\newcommand{\tb}{\theta}
\newcommand{\distance}{\mathbf{d}}
\newcommand{\open}{o}
\newcommand{\payment}{p}
\newcommand{\bid}{\mathbf{b}}
%UFL problem notations
\newcommand{\user}{u}
\newcommand{\facility}{\ell}
\newcommand{\opt}{texttt{OPT}}
\newcommand{\cost}{c}
\newcommand{\ocost}{c_o}
\newcommand{\scost}{c_s}
\newcommand{\frugal}{F}

\newcommand{\mechname}{\textsc{PredictedLimits}}
\newcommand{\mecherr}{\textsc{ErrorTolerant}}

\maketitle

\begin{abstract}

We study the problem of designing procurement auctions for the strategic uncapacitated facility location problem: a company needs to procure a set of facility locations in order to serve its customers and each facility location is owned by a strategic agent. Each owner has a private cost for providing access to their facility (e.g., renting it or selling it to the company) and needs to be compensated accordingly. The goal is to design truthful auctions that decide which facilities the company should procure and how much to pay the corresponding owners, aiming to minimize the total cost, i.e., the monetary cost paid to the owners and the connection cost suffered by the customers (their distance to the nearest facility). We evaluate the performance of these auctions using the \emph{frugality ratio}.

We first analyze the performance of the classic VCG auction in this context and prove that its frugality ratio is exactly $3$. We then leverage the learning-augmented framework and design auctions that are augmented with predictions regarding the owners' private costs. Specifically, we propose a family of learning-augmented auctions that achieve significant payment reductions when the predictions are accurate, leading to much better frugality ratios. At the same time, we demonstrate that these auctions remain robust even if the predictions are arbitrarily inaccurate, and maintain reasonable frugality ratios even under adversarially chosen predictions. We finally provide a family of ``error-tolerant'' auctions that maintain improved frugality ratios even if the predictions are only approximately accurate, and we provide upper bounds on their frugality ratio as a function of the prediction error.
\end{abstract}

\section{Introduction}
When government agencies, retail chains, or banking institutions need to open new facilities or branches to better serve their customers, they face a challenging optimization problem: they want customers to be as close as possible to a facility, but opening new facilities can be very costly, and the cost can vary significantly by location. Deciding which locations to open therefore requires balancing the customers’ \textit{connection costs} (their distance to the nearest facility) against the facilities' \textit{opening costs}. This optimization problem is known as the \emph{uncapacitated facility location} (UFL) problem and it has received a lot of attention in prior work (e.g., \cite{li2013approximating, guha1999greedy, shmoys1997approximation, CS03, mahdian2002improved, byrka2010optimal, li2013approximating}).

However, the vast majority of this prior work assumes that the designer has full information regarding both the connection and the opening costs, which is often unrealistic in practice. For example, each location may be owned by a different strategic agent who would need to be appropriately compensated for the agency to open a facility there. Negotiating this compensation (i.e., the opening cost) is itself a strategic problem: the owner wishes to maximize it, while the agency wishes to minimize it. This additional obstacle is captured by the \emph{strategic} UFL problem, where the opening cost of each location is private information held by the agent who own it~\cite{AT01, talwar2003price, CLWWZZ22, LiWZ22}.

From the agency’s perspective, rather than negotiating compensation separately with each owner, a more effective approach is to run a procurement auction. This encourages competition, leading to more efficient and cost-effective outcomes, which is one primary reason why the US government routinely uses procurement auctions~\cite{ProcurementGov}. To prevent manipulation, prior work on strategic UFL has focused on designing \emph{truthful} auctions that incentivize agents to reveal their true opening costs~\cite{AT01, talwar2003price, CLWWZZ22, LiWZ22}. Our work continues in this direction.

Evaluating the performance of a procurement auction is complicated by the lack of a suitable benchmark. If there is little competition, the agency may have to pay higher compensation, whereas greater competition can reduce costs. The \emph{frugality} literature~\cite{archer2002frugal, talwar2003price} addresses this by proposing the cost of the “second-best” solution as a benchmark (see Section~\ref{sec:prelims}). When this second-best cost is much higher than the optimum, the instance resembles a monopoly, and the agency must pay more. Conversely, if the second-best is close to the optimum, a well-designed auction can leverage competition for better outcomes. The \emph{frugality ratio} quantifies how closely the auction’s cost approaches this benchmark, and our main technical results are new frugality ratio bounds.

Despite its appeal, most prior work in frugal mechanism design is restricted to the adversarial framework, analyzing the auctions' performance under worst-case instances and assuming they have no information regarding the private costs. While this provides robust guarantees, it can be overly pessimistic, especially in practice, where agencies often have access to predictions regarding opening costs, derived from historical data, expert estimates, or data-driven models. This naturally raises the question: can we design mechanisms that exploit such predictions to improve frugality, while still remaining robust if the predictions turn out to be inaccurate?

Motivated by this, our work leverages the \textit{learning-augmented framework}, which has recently spurred significant research on ``algorithms with predictions''~\cite{mitzenmacher_vassilvitskii_2021,alps}, and more recently, on the design of truthful auctions and mechanisms augmented with predictions~
\cite{ABGOT22,XL22}. In this framework, the designer is given an unreliable prediction (potentially from machine learning or historical data), and the goal is to achieve improved guarantees when the prediction is accurate (\emph{consistency}), while maintaining strong guarantees even when the prediction is arbitrarily inaccurate (\emph{robustness}).
\subsection{Our Results}

Our first result is a tight analysis of the frugality ratio of the classic truthful VCG auction (without predictions) for the strategic UFL problem. Prior to this result, the best known upper bound was $4$ by \citet{talwar2003price}, and there were no known lower bounds. Using a tighter analysis and a matching lower bound, we prove that the frugality ratio of VCG is exactly $3$.

Then, rather than assuming that the auction has no information regarding the private opening cost $o_\ell$ at each location $\ell$, we consider \textit{learning-augmented auctions} that are provided with a prediction $\hat{o}_\ell$ regarding this cost. Crucially, this prediction is unreliable and can be arbitrarily inaccurate.
%Motivated by the potential to improve economic outcomes using external information, and inspired by recent trends in learning-augmented algorithms, we then investigate how potentially imperfect predictions of facility opening costs can be leveraged. We propose and analyze a novel mechanism that incorporates such predictions. We introduce a prediction-augmented mechanism that achieves near-optimal \emph{consistency}: when predictions perfectly match the true opening costs, its frugality ratio approaches 1 (specifically, $1+\epsilon$ for any desired $\epsilon > 0$).

Our second result is a new truthful procurement auction that takes a parameter $\epsilon\in (0, 2]$ as input and guarantees a frugality ratio of $1+\epsilon$ whenever the prediction is accurate (the \emph{consistency} guarantee), while simultaneously guaranteeing a frugality ratio of at most $\max\left\{5, 3 + \frac{2}{\epsilon}\right\}$, irrespective of how inaccurate the prediction may be (the \emph{robustness} guarantee). Note that choosing a small constant $\epsilon$ allows us to guarantee a near-optimal frugality (arbitrarily close to $1$) when the prediction is accurate, while simultaneously guaranteeing a constant frugality even for adversarially chosen predictions.
To achieve this guarantee, our learning-augmented  auction uses the VCG mechanism on input that is scaled using the predictions. Specifically, if the opening cost reported by the owner at some location $\ell$ is exceeds the prediction $\hat{o}_\ell$, then it is scaled up even higher, reducing their chance of being selected. When the prediction is accurate, this limits the agents' ability to manipulate, leading to reduced costs.
%achieves this guarantee is by treating the agent at each locations, $\ell$, differently depending on whether they request a payment above or below $\hat{o}_\ell$. Specifically, if it is above, then the requested payment is scaled up even further, reducing the chances that this location will be served. This, effectively, penalizes the bidder and incentivizes them to remain below the predicted cost if they can (which is what they do if the prediction is accurate).
%Crucially, we demonstrate that our mechanism maintains strong \emph{robustness} guarantees. Even when faced with arbitrarily inaccurate, adversarial predictions, the mechanism's frugality ratio is bounded by $\max\left(5, 3 + \frac{2}{\epsilon}\right)$
%These results highlight the potential of integrating machine-learned predictions into mechanism design to achieve significantly more frugal outcomes in strategic settings like the UFL problem, without sacrificing worst-case performance guarantees.

For our third result, we design a learning-augmented procurement auction that is ``error tolerant.'' We first define the prediction error as $\eta = \max_{\ell \in L} \max\left\{\hat{o}_\ell/o_\ell, o_\ell /\hat{o}_\ell\right\}$, i.e., the largest ratio between the predicted opening cost and the actual opening cost. Our auction is then provided with an error tolerance parameter, $\lambda>1$, as input (as well as the parameter $\epsilon\in (0, 2]$), and as long as the error $\eta$ of the prediction is at most $\lambda$, it guarantees a frugality ratio of $\eta(1+\lambda)+2\epsilon$. Even if the error exceeds the error tolerance threshold, i.e., $\eta > \lambda$, the frugality ratio is at most $\max\left\{\,2\lambda^4+3\lambda^2,\;3+\tfrac{2}{\epsilon}\right\}$.

\subsection{Related Work}

%\xnote{reorg the paragraph}

%\vnote{I think maybe the remaining three paragraphs make sense, with the first one focusing on the specific problem of strategic UFL and only keeping the particularly relevant references for the non-strategic UFL. Also, in the third paragraph I would maybe focus on mechanism design with predictions and cite just a few alg papers}

%\paragraph{Uncapacitated Facility Location}
The strategic UFL problem was first studied by \citet{AT01}, who analyzed the structure of truthful mechanisms for this problem. The frugality of truthful mechanisms for the strategic UFL problem was studied by \citet{talwar2003price}, who showed that the VCG auction has a frugality ratio of at most $4$. \citet{CLWWZZ22} considered different social cost objectives for the special case where the facility locations coincide with the locations of the agents (i.e., each agent has a ``dual-role'' as a customer or facility operator). \citet{CLWWZZ22} and \citet{LiWZ22} also considered the budget-feasible version of this problem, where the overall payment cannot exceed a budget. For related work on the non-strategic version of the UFL problem, see Appendix~\ref{app:relatedwork}.

%The focus of all this work was on the design of truthful mechanisms aiming to minimize the social cost. 
%\vnote{Compress rest of paragraph a lot or drop? defer to appendix? I would avoid the discussion regarding running time, since we use VCG}
%The metric Uncapacitated Facility Location (UFL) problem is a fundamental NP-hard optimization problem, widely studied due to its theoretical significance and practical relevance in operations research and algorithmic game theory. Early foundational results established the APX-hardness of the problem, showing that it cannot be approximated within a factor better than $1.463$ unless $P=NP$ \cite{guha1999greedy}. The first constant-factor approximation was provided by \citet{shmoys1997approximation}, who achieved a ratio of approximately $3.16$ . Subsequent improvements significantly narrowed the approximation gap, with notable progress by \citet{CS03} to 1.736 and \citet{mahdian2002improved} to $1.52$-approximation via sophisticated linear-programming techniques. \citet{byrka2010optimal} later improved this bound to a $1.50$-approximation using a novel bifactor algorithm . The current state-of-the-art algorithm, due to \citet{li2013approximating}, achieves an approximation ratio of $1.488$, leaving a small gap relative to the known hardness bound. 
% These milestones collectively underscore the theoretical depth of the UFL problem and highlight the subtle complexity involved in narrowing the approximation gap further.

%\paragraph{Frugal Mechanism Design.}
The study of frugal mechanism design is motivated by the challenge of minimizing unnecessary overpayment while preserving truthfulness in procurement auctions and combinatorial team selection problems. This perspective has been applied across a variety of classic optimization settings. In network design, \citet{archer2002frugal} and \citet{talwar2003price} initiated the investigation for path auctions, demonstrating that VCG mechanisms can incur large overpayments and establishing tight frugality bounds under the frugal solution benchmark. For coverage problems such as vertex cover and set cover, \citet{elkind2007frugality} and subsequent works developed truthful mechanisms with frugality ratios that depend on structural parameters like graph degree or spectral properties, and also provided matching lower bounds in several cases. 
Frugality in $k$-flow and cut problems has also been explored, with mechanisms achieving constant-competitive frugality ratios relative to various benchmarks, including both disjoint-alternative and equilibrium-based definitions~\cite{karlin2005beyond, kempe2010spectral, chen2010frugal}. In some settings, such as matroid and spanning tree auctions, VCG mechanisms are provably frugal, achieving optimal or near-optimal frugality ratios~\cite{karlin2005beyond,talwar2003price}. 

The design of learning-augmented mechanisms for settings involving strategic agents was initiated by \citet{ABGOT22} and \citet{XL22}. This line of work spans strategic facility location \cite{ABGOT22, XL22, IB22, BGT24, chen2024strategic, balkanski2024randomized,shi2025prediction}, strategic scheduling \cite{XL22, BGT223, CSV24}, auction design \cite{MV17, XL22, LuWanZhang23, caragiannis2024randomized, BGTZ23, caragiannis2025mechanisms,GST25}, bicriteria mechanism design for welfare and revenue trade‑offs \cite{BPS23}, graph problems with private input \cite{CKST24}, distortion \cite{BFGT23, FKV25}, and equilibrium analysis \cite{GKST22, IBB24}. 
%metric distortion guarantees \cite{BFGT23}, and equilibrium analyses \cite{GKST22, IBB24}. 
% More recently, \cite{} shifted the focus from input predictions to outcome‑space forecasts. \vnote{Too much emphasis on this last paper, no? In general, we should have a more consistent use of cite vs citet and what we focus on should be well motivated}
The work most closely related to this paper is \cite{XL22}, which studied frugal mechanism design with predictions in the context of path auctions. See Appendix~\ref{app:relatedwork} for a more extensive discussion on the line of work on algorithms with predictions.
 %For a broader perspective on this rapidly evolving field, see \cite{BGT23}. \xnote{add the new ones}

% \newpage
\section{Preliminaries}\label{sec:prelims}
%\paragraph{UFL.} 
In the \emph{Uncapacitated Facility Location (UFL) problem}, there is  a set \( U \) of users and a set \( L \) of facilities. Each facility \( \ell \in L \) has an opening cost \( o_\ell \). Each user $u \in U$ and facility $\ell \in L$ have a connection cost $d(u, \ell)$. The connection costs are assumed to form a metric space, i.e., $d(x,y)$ is defined for any $x, y \in U \cup L$ and satisfies $d(x,x) = 0$, $d(x,y) \geq 0$ (non-negativity), $d(x, y) = d(y,x)$ (symmetry), and $d(x,z) \leq d(x,y) + d(y, z)$ (triangle inequality) for all $x,y,z \in U \cup L$. The cost of connecting a user $u \in U$ to a set of facilities $S \subseteq L$ is $\min_{\ell \in S} d(u, \ell)$, i.e., it is its distance to  its closest facility in $S$. Given facilities $S \subseteq L$, we say  that a user $u$ is assigned to facility $\ell$ if $\ell = \argmin_{\ell \in S} d(u, \ell)$. The total connection cost incurred by a set of  facilities $S \subseteq L$ is
\[d(U,S) = \sum_{u \in U} \min_{\ell \in S} d(u, \ell),\]
and its total cost $c(S)$  is the sum of its opening cost  and  total connection cost, i.e.,
\[
c(S) = \sum_{\ell \in S} o_\ell + d(U,S).
\]
The goal is to open a set of facilities $S \subseteq L$ that minimizes the total cost. Given $U, d$, and $\mathbf{o}=(o_\ell)_{\ell\in L}$, the optimal facility set is $\OPT(U, \mathbf{o}, d) = \argmin_{S \subseteq L} c(S)$.

%goal is to open a set facilities $S \subseteq U$ that minimizes the total cost, which is the sum of the opening cost and connection costs. 

%Each facility \( \ell \in L \) has an opening cost \( o_\ell \), and there is a public connection cost \( d(u, \ell) \) for connecting a user \( u \in U \) to a facility \( \ell \). The costs satisfy the standard metric properties: symmetry \( d(u, \ell) = d(\ell, u) \) and the triangle inequality. For presentation purposes, we denote 
%\[d(U,S) = \sum_{u \in U} \min_{\ell \in S} d(u, \ell).\]
%i.e., the distance between a set of users to a set of facilities are the total distance of the users to it's closest facility in $S$. Given a set of open facilities \( S \subseteq L \), the total \emph{cost} is defined as:
%\[
%c(S) = \sum_{\ell \in S} o_\ell + \sum_{u \in U} d(U,S),
%\]
%where the first term is the total opening cost and the second term is the total service cost, i.e., the cost of connecting each user to their closest open facility. The objective is to minimize the total cost.

\paragraph{Strategic UFL.} In the strategic version of the  UFL problem, each facility \( \ell \in L \) is owned by a strategic agent and its opening cost \( o_\ell \) is private  (the connection costs are public information). An auction $\mech(U, \mathbf{b}, d)$ for strategic UFL takes as input the set of users $U$ and the connection costs $d$, and it asks the owner of each location $\ell\in L$ to report an opening cost $b_{\ell}$ (which we refer to as a bid), leading to a bid vector $\mathbf{b}=(b_\ell)_{\ell\in L}$. For simplicity, we write $\mech(\mathbf{b})$ and $\OPT(\mathbf{o})$ when $U$ and $d$ are clear from context. We also use $\mathbf{b}_{-\ell}$ to refer to the vector of all bids excluding $b_\ell$. 
%write $(b, \mathbf{b}_{-\ell})$ as the  bid profile such that $b$ is the bid of $\ell$ and $b_{\ell'}$ is the bid of $\ell' \neq \ell$.  
We say that a facility $\ell$ misreports if its reported cost is not its true cost, i.e., $b_\ell \neq o_\ell$. The output $(S, \mathbf{p})$ of an auction consists of a set $S \subseteq L$ of facilities to open and a payment $p_\ell$ to each facility $\ell \in S$ for opening. The utility of a facility from output $(S, \mathbf{p})$ is 
$$u_\ell(S, \mathbf{p}) = \begin{cases} p_\ell - o_\ell & \text{ if } \ell \in S \\ 0 & \text{ otherwise} \end{cases}.$$
An auction is \emph{truthful} if it is a dominant strategy for every facility $\ell$ to report its true opening cost, i.e., $b_\ell = o_\ell$, irrespective of what the other facilities report. That is, for every $o_\ell, b_\ell, \mathbf{b}_{-\ell}, U, d$,
\[
u_\ell(\mech(o_\ell, \mathbf{b}_{-\ell})) \geq u_\ell(\mech(b_\ell, \mathbf{b}_{-\ell})).
\]
%Each agent reports a bid \( b_\ell \), and a mechanism \( \mech(U, \mathbf{b}) \to (S, \mathbf{p}) \) selects a subset \( S \subseteq L \) of facilities to open and assigns payments \( p_\ell \) to facilities in \( S \). The utility of a selected facility \( \ell \in S \) is:
%\[
%u_\ell(U, \mathbf{b}, d) = p_\ell - o_\ell,
%\]
%and the utility of an unselected facility is zero. 
% \paragraph{Monotonicity and Myerson’s Lemma.}
An auction \( \mech \) for UFL is \emph{monotone} if for any facility \( \ell \in L \) and any bid profile \( \mathbf{b}_{-\ell} \) of the other facilities, the probability that $\ell \in S$, where $S$ is the facilities chosen by $\mathcal{M}(b_\ell , \mathbf{b}_{-\ell})$, is non-increasing in its bid \( b_\ell\). In single-parameter environments such as strategic UFL, Myerson’s Lemma characterizes truthful auctions:

\begin{lemma}[Myerson’s Lemma]\label{lem:myerson}
An auction is truthful if and only if it is monotone. For any monotone auction, there exists a unique payment rule that ensures truthfulness, which can be computed explicitly.
\end{lemma}
\paragraph{Frugality and Efficiency.} The total cost incurred by an auction for outputting   $(S, \mathbf{p})$ is the sum of the payments and the total connection cost, i.e.,
\[
p(S, \mathbf{p}) = \sum_{\ell \in S} p_\ell + d(U,S).
\]
The \emph{frugal facility set} is the second-best solution, i.e.,
\[
\frugal(U, \mathbf{o},d) = \argmin_{S \subseteq L \setminus \OPT(U, \mathbf{o},d)} c(S).
\]
The efficiency of an auction is evaluated by its \emph{frugality ratio}, which is the worst-case ratio between its total cost and the cost of the frugal solution:
\[
\texttt{frugality}(\mech) = \max_{U, \mathbf{o},d} \frac{p(\mech(U, \mathbf{o},d))}{c(\frugal(U, \mathbf{o},d))}.
\]
%Given an instance \( I = (U, \mathbf{o}, d) \), 
% \enote{an instance also depends on the connection costs $d$}
%the \emph{optimal facility set} is:
%\[
%\OPT(U, \mathbf{o},d) = \argmin_{S \subseteq L} c(S),
%\]
%and the \emph{frugal facility set} is the optimal solution after removing all facilities in the optimal solution:
%\[
%\frugal(U, \mathbf{o},d) = \argmin_{S \subseteq L \setminus \OPT(U, \mathbf{o})} c(S).
%\]
\paragraph{Learning-Augmented Framework.} In the \emph{learning-augmented setting}, the auction is  given predictions \( \hat{\mathbf{o}} \) about the opening costs $\mathbf{o}$. %Given an instance \( (U, d, \mathbf{o}, \hat{\mathbf{o}}) \), we denote the total cost of the auction as \( p(\mech(U,  \mathbf{o}, d,  \hat{\mathbf{o}})) \). 
We evaluate the performance of \( \mech \) using two metrics.

\textbf{Consistency:} The frugality ratio when the predictions are accurate (\( \hat{\mathbf{o}} = \mathbf{o} \)):
    \[
    \texttt{consistency}(\mech) = \max_{U, \mathbf{o}, d} \frac{p(\mech(U, \mathbf{o},  d, \hat{\mathbf{o}} = \mathbf{o} ))}{c(F(U, \mathbf{o}, d))}.
    \]

\textbf{Robustness:} The frugality ratio when the predictions can be arbitrarily wrong:
    \[
    \texttt{robustness}(\mech) = \max_{U, \mathbf{o}, d, \hat{\mathbf{o}}} \frac{p(\mech(U, \mathbf{o}, d, \hat{\mathbf{o}}))}{c(F(U, \mathbf{o}))}.
    \]

%%%%%%%%%%%%%%%%%%%%%%%%%%%%

\section{Tight Frugality Bounds for the Vickrey-Clarke-Groves Auction}\label{sec:VCG-UFL}
% \xnote{add discussion here that most frugal mechanism design literature studies VCG, and no mechanism better than VCG is know in any problems (this is only true for the benchmark we consider tho, for the other benchmarks there are other mechanisms that outperform VCG)}

% \vnote{Do we need to discuss the Talwar result again?}
% \citet{talwar2003price} provided a frugality upper bound of VCG at $4$.

% \xnote{I dont know what else to say here other than 
%   } \vnote{Just say that and save some space}
Our first main result establishes an improved frugality ratio of \(3\) for the VCG auction, { which we start by describing in the context of the strategic UFL problem.}

\paragraph{The VCG auction.} Given an instance \((U,\mathbf{o},d)\) of UFL, the set of facilities opened by the VCG auction is the optimal facility set \(\OPT(\mathbf{o})\). The VCG payment, also called threshold payment, to each opened facility \(\ell \in \OPT(\mathbf{o})\) is
\[
p_\ell
\;=\;
c\bigl(\OPT(\infty, o_{-\ell})\bigr)
\;-\;
c\bigl(\OPT(0,\mathbf o_{-\ell})\bigr),
\]
i.e., it is the difference between the total cost of the optimal solution when facility \(\ell\) is infeasible (\(o_\ell=\infty\)) and when \(\ell\) is free (\(o_\ell=0\)). This payment can be alternatively defined as
\[
p_\ell
\;=\;
\max\bigl\{\,b\ge0 : \ell\in\OPT(b,\mathbf o_{-\ell})\bigr\},
\]
i.e., the largest bid that facility \(\ell\) can declare and remain in the optimal set \(\OPT\) selected by VCG.

\paragraph{Analysis of VCG.} We first give an upper bound of \(3\) on the frugality ratio of VCG.
At a high level, the proof bounds the total cost incurred by the VCG auction by (1) bounding the payments by twice the total cost of the frugal solution and (2) observing that the total connection cost of the optimal solution is bounded by its own total cost, which is in turn bounded by the total cost of the frugal solution. Combining these two facts immediately yields a \(3\)-approximation. The payment bound is achieved by a ``rerouting'' argument: for each facility in the winning set, we show how to reassign its users to other facilities, either within the winning set itself or to those in some other set (the frugal set for example), thereby upper-bounding its threshold payment by the corresponding rerouting cost. 

We first provide the payment bound via the rerouting argument, we state it in a more general setting since it will also be used for the analysis of the mechanism with predictions.

\begin{lemma}[Unified payment bound under set-dependent scaling]\label{lem:unified-wprime}
Assume a mechanism outputs $W\subseteq L$ such that 
\[
W \in \argmin_{S \subseteq L} \left(d(U,S)\ +\ \sum_{g\in S}\alpha_g(S)\,o_g\right),
\]
where $\alpha_g(S)\ge 0$ are parameters chosen by the mechanism.
For $\ell\in W$, let $p_\ell$ be its threshold payment and let $U_\ell$ be the users assigned to $\ell$ under $W$.
Fix any subset $W'\subseteq W$ and any reference set $R\subseteq L$.
Define $\alpha_f^{\max}:=\sup\{\alpha_f(S): f\in S\subseteq L\}$.
Then
\[
\sum_{\ell\in W'} \alpha_\ell(W)\,p_\ell\ \le\ \sum_{f\in R}\alpha_f^{\max}\,o_f\ +\ 2\,d(U,R).
\]
\end{lemma}

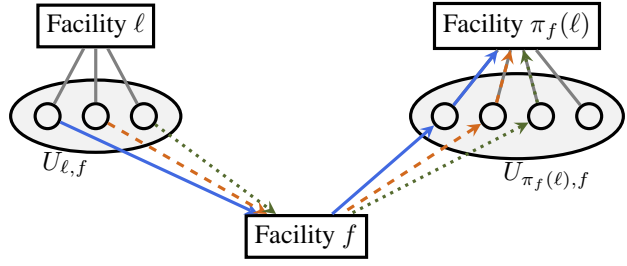
\begin{figure}
    \centering
    \begin{tikzpicture}[auto, line width=1.2pt, scale=0.8]
      % Left cluster: facility ℓ and clients U_{ℓ,f}
      \node[draw, rectangle, minimum width=1.5cm] (li) at (0,0) {Facility $\ell$};
      \begin{scope}[shift={(0,-1.5)}]
        \node[draw, ellipse, fill=gray!10, inner xsep=0.8cm, inner ysep=0.34cm] (Ui) {};
        \node[draw, circle] (ui1) at (-0.8,0) {};
        \node[draw, circle] (ui2) at (0,0) {};
        \node[draw, circle] (ui3) at (0.8,0) {};
        \foreach \x in {ui1,ui2,ui3} {
          \draw[gray] (\x) -- (li);
        }
      \end{scope}
      \node at (-0.5,-2.3) {$U_{\ell,f}$};

      % Right cluster: facility π_f(ℓ) and clients U_{π_f(ℓ),f}
      \node[draw, rectangle, minimum width=1.5cm] (li1) at (7,0) {Facility $\pi_f(\ell)$};
      \begin{scope}[shift={(7,-1.5)}]
        \node[draw, ellipse, fill=gray!10, inner xsep=1cm, inner ysep=0.4cm] (Uk) {};
        \node[draw, circle] (uk1) at (-1.2,0) {};
        \node[draw, circle] (uk2) at (-0.4,0) {};
        \node[draw, circle] (uk3) at (0.4,0) {};
        \node[draw, circle] (uk4) at (1.2,0) {};
        \foreach \x in {uk1,uk2,uk3,uk4} {
          \draw[gray] (\x) -- (li1);
        }
      \end{scope}
      \node at (7.5,-2.5) {$U_{\pi_f(\ell),f}$};

      % Bottom facility f
      \node[draw, rectangle, minimum width=1.5cm] (fj) at (3.5,-3.5) {Facility $f$};

      % Original connections
      \draw[RoyalBlue, solid, -stealth]  (uk1) -- (li1);
      \draw[Chocolate, dashed, -stealth] (uk2) -- (li1);
      \draw[DarkOliveGreen, dotted, -stealth] (uk3) -- (li1);

      \draw[RoyalBlue, solid, -stealth]  (ui1) -- (fj);
      \draw[Chocolate, dashed, -stealth] (ui2) -- (fj);
      \draw[DarkOliveGreen, dotted, -stealth] (ui3) -- (fj);

      % Rerouting arrows
      \draw[->, RoyalBlue, solid, stealth-]      (uk1) -- (fj);
      \draw[->, Chocolate, dashed, stealth-]      (uk2) -- (fj);
      \draw[->, DarkOliveGreen, dotted, stealth-] (uk3) -- (fj);
    \end{tikzpicture}
    \caption{Illustration of how users in $U_{\ell,f}$ are rerouted to facility $\pi_f(\ell)$. Each color (and line pattern) denotes the path taken by a distinct user $u \in U_{\ell,f}$. The total connection‐cost along each set of colored edges gives an upper bound on the payment to facility $\ell$ for serving that subset of users.}
    \label{fig:rerouting-wp}
\end{figure}

\begin{proof}[Proof of Lemma~\ref{lem:unified-wprime}]
Let $W$ and $R$ be the chosen and reference sets, respectively.
We denote by $U_\ell \subseteq U$ the users assigned to facility $\ell$ according to $W$.
We first note that for any set of facilities $O_\ell \subseteq L \setminus \{\ell\}$ that does not contain facility $\ell$, we must have
\begin{equation}
\label{eq:one-wp}
  \alpha_\ell(W)\,p_\ell + \sum_{u \in U_\ell} d(u, \ell)
  \;\le\;
  \sum_{f \in O_\ell \setminus W} \alpha_f(O_\ell)\,o_f + \sum_{u \in U_\ell} \min_{v \in O_\ell} d(u, v),
\end{equation}
otherwise $O_\ell$ would have a smaller total $C'$-cost than $W$ when facility $\ell$ reports opening cost $p_\ell$, which would imply
\[
p_\ell>\max\bigl\{\,b\ge0 : \ell\in W(b,\mathbf o_{-\ell})\bigr\},
\]
and would contradict the alternative definition of the threshold payment.

The central part of the proof consists of defining some alternative solution $O_\ell$ and to reassign and reroute the users in $U_\ell$ to facilities in $O_\ell$ to bound $ \sum_{u \in U_\ell} \min_{v \in O_\ell} d(u, v)$.
To define $O_\ell$ and this reassignment, we first need to introduce some notation.
For each pair \( (\ell \in W, f \in R) \), we let  \( U_{\ell,f} \subseteq U \) be the set of users assigned to \( \ell \) in the chosen solution $W$ and to \( f \) in the reference set $R$.

Crucially, we define our groupings with respect to the subset $W'$. For each \( f \in R \), let
\[
  W'_f = \{\ell \in W' : |U_{\ell,f}| > 0\}
\]
and let $x_f(\ell)$ be the ranking of $\ell \in W'_f$ among all other facilities in $W'_f$
in terms of the size of $|U_{\ell,f}|$, where ties are broken arbitrarily but consistently.
This definition implies that $x^{-1}_f(j) \in W'_f$ is the facility $\ell$ with the $j^{\text{th}}$ smallest non-negative  $|U_{\ell,f}|$ and we thus have
$
  0 < |U_{x^{-1}_f(1),f}| \le \dots \le |U_{x^{-1}_f(|W'_f|),f}|.
$

We define the alternate solution to be
\[
O_\ell \;=\; W \setminus \{\ell\} \ \cup\ \{\,f \in R:\ x_f(\ell) = |W'_f|\,\}.
\]
In this solution, facility $\ell$ is removed from $W$ and, for each $f\in R$, if $|U_{\ell, f}|$ is largest among the facilities in the subset $W'_f$, the facility $f$ is added.
We then define a mapping
$
  \pi_f: W'_f \to W'_f \cup \{f\}
$
that reassigns users in $U_{\ell,f}$ that are assigned to facility $\ell$ according to $W$ to a new facility in $W'_f \cup \{f\}$.
This reassignment is defined as
\[
  \pi_f(\ell)=
  \begin{cases}
  x^{-1}_f (x_f(\ell)+1) & \text{ if } x_f(\ell) < |W'_f|,\\[2pt]
  f & \text{ if } x_f(\ell) = |W'_f|.
  \end{cases}
\]
In words, \(\pi_f(\ell)\) maps facility \(\ell\) to the facility ranked immediately after $\ell$ with respect to \(f\); if \(\ell\) is ranked last, then \(\pi_f(\ell)=f\).
Since  $\bigcup_{f : \ell \in W'_f} U_{\ell, f} = \bigcup_{\ell \in W'} U_{\ell,f}$, we have that for any $\ell \in W'$,
\begin{align}
\label{eq:two-wp}
    \sum_{u \in U_\ell} \min_{v \in O_\ell} d(u, v)
    \ \leq\ \sum_{f : \ell \in W'_f} \min_{v \in O_\ell} d(U_{\ell, f}, v)
    \ \leq\ \sum_{f : \ell \in W'_f}  d\big(U_{\ell, f}, \pi_f(\ell)\big).
\end{align}

Next, if $x_f(\ell) < |W'_f| ,$ then $|U_{\ell, f}| \leq |U_{\pi_f(\ell), f}|$ by the definition of $\pi_f$.
Thus, combined with the triangle inequality, we get that  if $x_f(\ell) < |W'_f| ,$ then
\begin{align}
\label{eq:three-wp}
d\big(U_{\ell, f}, \pi_f(\ell)\big)
\ \leq\
d\big(U_{\ell,f}, f\big)
+ d\big(f, U_{\pi_f(\ell),f}\big)
+ d\big(U_{\pi_f(\ell),f}, \pi_f(\ell)\big).
\end{align}
Informally, this last inequality corresponds to rerouting the users in $U_{\ell, f}$ to $\pi_f(\ell)$ by going through $f$ and then $U_{\pi_f(\ell),f}$.
This rerouting is the key component of this proof, see Figure~\ref{fig:rerouting-wp} for an illustration.
We obtain that, summing over \(\ell\in W'\),
{\allowdisplaybreaks
\begin{align*}
 &\sum_{\ell \in W'}\!\!\left(\alpha_\ell(W)\,p_\ell + \sum_{u \in U_\ell} d(u, \ell)\right)\\
 & \quad \leq \sum_{\ell \in W'}\!\!\left(\sum_{f: x_f(\ell) = |W'_f|} \alpha_f(O_\ell)\,o_f \ +\ \sum_{u \in U_\ell} \min_{v \in O_\ell} d(u, v)\right)
 && \text{by }\eqref{eq:one-wp}\\
 &\quad\leq \sum_{\ell \in W'}\!\!\left(\sum_{f: x_f(\ell) = |W'_f|} \alpha_f(O_\ell)\,o_f \ +\ \sum_{f : \ell \in W'_f}  d\big(U_{\ell, f}, \pi_f(\ell)\big) \right)
 && \text{by }\eqref{eq:two-wp}\\
 &\quad\leq \sum_{f \in R}\!\left(\alpha_f^{\max} o_f \ +\ \sum_{\ell \in W'_f}  d\big(U_{\ell, f}, \pi_f(\ell)\big) \right)
 && \text{(Summing by } f)
\\
 &\quad= \sum_{f \in R}\!\left(\alpha_f^{\max} o_f \ +\ d\!\left(U_{x_f^{-1}(|W'_f|), f},f\right)
        + \sum_{j=1}^{|W'_f|-1}  d\!\left(U_{x_f^{-1}(j), f}, x_f^{-1}(j+1)\right) \right)
\\
 &\quad\leq \sum_{f \in R}\!\Bigg(\alpha_f^{\max} o_f \ +\ d\!\left(U_{x_f^{-1}(|W'_f|), f},f\right)
\\[-2pt]
 &\quad \quad
        + \sum_{j=1}^{|W'_f|-1}  \big( d(U_{x_f^{-1}(j),f},f)
       + d(f, U_{x_f^{-1}(j+1),f})
       + d(U_{x_f^{-1}(j+1),f},x_f^{-1}(j+1))\big) \Bigg)
 && \text{by }\eqref{eq:three-wp}\\
 &\quad\leq \sum_{f \in R}\!\left(\alpha_f^{\max} o_f + \sum_{\ell\in W'_f}\big( d(U_{\ell,f},f)
       + d(f, U_{\ell,f})
       + d(U_{\ell,f},\ell)\big) \right)\\
 &\quad= \left(\sum_{f \in R} \alpha_f^{\max} o_f + 2\sum_{f \in R} \sum_{\ell\in W'_f} d(U_{\ell,f},f)\right) + \sum_{f \in R} \sum_{\ell \in W'_f}
        d(U_{\ell,f},\ell).
\end{align*}}
We observe that the last term is exactly the connection cost of the subset $W'$:
\[
\sum_{f \in R} \sum_{\ell \in W'_f} d(U_{\ell,f},\ell) = \sum_{\ell \in W'} \sum_{f : \ell \in W'_f} d(U_{\ell,f}, \ell) = \sum_{\ell \in W'} \sum_{u \in U_\ell} d(u, \ell).
\]
This cancels perfectly with the connection cost term on the LHS. Furthermore, $\sum_{\ell \in W'_f} d(U_{\ell,f}, f) \le d(U, f)$.
Thus, we conclude
\[
\sum_{\ell\in W'} \alpha_\ell(W)\,p_\ell
\ \le\
\sum_{f \in R} \alpha_f^{\max} o_f \ +\ 2\sum_{f \in R} d(U,f)
\ =\
\sum_{f \in R} \alpha_f^{\max} o_f \ +\ 2\,d(U,R).
\qedhere\]
\end{proof}

\begin{corollary}\label{cor:vcg_frugality}
    The frugality ratio of the VCG auction for the metric uncapacitated facility location problem is at most $3$.
\end{corollary}
\begin{proof}We apply Lemma~\ref{lem:unified-wprime} with the winner set $W = \OPT$, the subset $W' = \OPT$, and the reference set $R = F$. Since the VCG auction uses unscaled costs, the multipliers are $\alpha_\ell(S) = 1$ for all $\ell$ and $S$, implying $\alpha_f^{\max} = 1$. Substituting these values into the lemma yields:$$\sum_{\ell\in\OPT} p_\ell \ \le\ \sum_{f\in F} o_f \;+\; 2\,d(U,F).$$Noting that $\sum_{f \in F} o_f + 2d(U,F) = c(F) + d(U,F)$ and that the connection cost satisfies $d(U,\OPT) \le c(\OPT) \le c(F)$, we sum the payments and connection cost:$$\sum_{\ell\in\OPT} p_\ell \;+\; d(U,\OPT)\ \le\ \bigl(c(F) + d(U,F)\bigr) + c(F) \ \le\ 3\,c(F).\qedhere$$\end{proof}
% \begin{proof}
% % (replacement via Lemma~\ref{lem:unified-wprime})]
% Apply Lemma~\ref{lem:unified-wprime} with $W'=W=\OPT$, $R=F$, and $\alpha_\ell(S)\equiv 1$ for any subset $S$ we get
% \[
% \sum_{\ell\in\OPT} p_\ell \ \le\ \sum_{f\in F} o_f \;+\; 2\,d(U,F).
% \]
% Since $d(U,\OPT)\le c(F)$, we conclude
% \[
% \sum_{\ell\in\OPT} p_\ell \;+\; d(U,\OPT)\ \le\ 3\,c(F).\qedhere
% \]
% \end{proof}

Next, we show a lower bound on the frugality of VCG, which holds even for tree metrics. 
% Due to space limitations, we defer its analysis to Appendix~\ref{app:vcg}.}
% We now show that the frugality ratio of \(3\) for VCG is tight by exhibiting the following instance.
% provide a lower bound for the VCG auction, demonstrating that the worst-case frugality ratio can not be better than $3$ in the limit.
% \xnote{say it's three in the limit}
\begin{lemma}
    The frugality ratio of VCG auction is at least $3 - \frac{6}{|L|+1}$,  even for tree metrics. 
\end{lemma}

\begin{proof}
    Consider the tree metric instance with $k$ users $U = \{u_1, \ldots, u_{k}\}$ and $k+1$ facilities $L = \{\ell_0, \ell_1, \ldots, \ell_{k}\}$.
    The graph forms a star centered at $\ell_0$. 
    Each peripheral facility $\ell_i$ is connected to $\ell_0$ via a user $u_i$, such that the path is $\ell_0 - u_i - \ell_i$. 
    Distances are set to $d(\ell_0, u_i) = 1$ and $d(u_i, \ell_i) = 1$. 
    Distance fro any user $u_i$ to any peripheral facility $\ell_j$ where $j \neq i$ is $d(\ell_j, u_i) = 3$.
    The central facility $\ell_0$ has an opening cost $o_{\ell_0} = 2$, and each peripheral facilities $\ell_i$ ($i \ge 1$) have $o_{\ell_i} = 0$ 
    (see Figure~\ref{fig:bad_instance} for an illustration).
\begin{figure}[ht]
    \centering  % Replaces \begin{center} for better spacing
    \begin{tikzpicture}[scale = 0.8]
        % Parameters
        \def\facilityradius{3}
        \def\k{6}

        % Draw central facility
        \node[draw, fill=gray!30, rectangle ] (central) at (0,0) {\textbf{2}};

        % Draw peripheral facilities and users
        \foreach \i in {1,...,\k} {
            \pgfmathsetmacro{\angle}{360/\k * (\i - 1)}
            \node[draw, fill=gray!30, rectangle ] (facility\i) at (\angle:\facilityradius) {\textbf{0}};
            \pgfmathsetmacro{\userdist}{\facilityradius/2}
            \node[draw, circle, minimum size=8pt, inner sep=0pt] (user\i) at (\angle:\userdist) { };
            \draw[thick,  gray] (central) -- (user\i);
            \draw[thick, gray] (user\i) -- (facility\i);
        }

        % Add distance labels
        \foreach \i in {1,...,\k} {
            \pgfmathsetmacro{\angle}{360/\k * (\i - 1)}
            \node at (\angle-5:2.25) {\textcolor{blue}{\textbf{1}}};
            \node at (\angle-15:0.8) {\textcolor{blue}{\textbf{1}}};
        }
    \end{tikzpicture}
    \caption{Illustration of the lower bound instance where each square represents a facility location with its opening cost, and the circles represent users.}
    \label{fig:bad_instance} % <--- MOVED HERE (Inside scope, right after caption)
\end{figure}
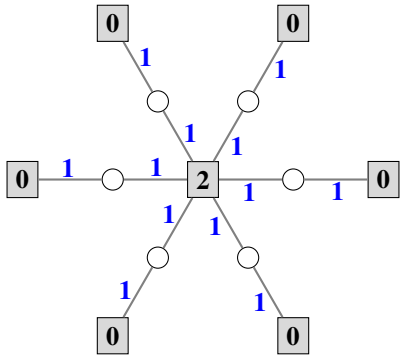

    The optimal solution $\OPT$ is to open all peripheral facilities $\{\ell_1, \ldots, \ell_k\}$. 
    Each user connects to its closest $\ell_i$ at cost $1$.$$c(\OPT) = \sum_{i=1}^k o_{\ell_i} + \sum_{i=1}^k d(u_i, \ell_i) = k \cdot 0 + k \cdot 1 = k.$$
    The only alternative (hence frugality) solution $F = \{\ell_0\}$ has a cost of $c(S_{\ell_0}) = o_{\ell_0} + \sum_{i=1}^k d(u_i, \ell_0) = 2 + k = k+2$.
    The VCG auction selects the optimal solution $\OPT = \{\ell_1, \ldots, \ell_k\}$.
    The VCG payment $p_{\ell_i}$ for each $\ell_i \in \OPT$ is:
    $$p_{\ell_i} = c(\OPT(\infty, \mathbf{\open})) - c(\OPT(0, \mathbf{\open}_{-i}))$$
     If $o_{\ell_i}=\infty$, the new optimal solution is $S_{\ell_0}$ with cost $k+2$, i.e.,
     $$c(\OPT(\infty, \mathbf{\open})) = k+2.$$
    Total cost when $\ell_i$ has a cost of $0$ is the original optimal cost $k$:
    $$c(\OPT(0, \mathbf{\open}_{-i})) = k.$$
    Therefore, $p_{\ell_i} = (k+2) - k = 2$.
    The total payment to the facilities is $\sum p_{\ell_i} = 2k$.
    The total VCG cost for the numerator is defined as the total payments plus the total user connection cost:$$p(\text{VCG}) = \sum_{i = 1}^{k} p_{\ell_i} + \sum_{i = 1}^k d(u_i, \ell_i) = 2k + k = 3k.$$
    Using the cost of the central facility solution as the baseline, $c(F(U, \open)) = k+2$, the frugality ratio is:$$\frac{c_{VCG}}{c(F(U, \open))} = \frac{3k}{k+2}.$$
    Since $k = |L|-1$, we substitute this into the ratio:
    \[\frac{3(|L| - 1)}{|L|+1} = \frac{3(|L| + 1) - 6}{|L|+1} = 3 - \frac{6}{|L|+1}. \qedhere\]
\end{proof}
The frugality ratio of the VCG auction follows from the two previous lemmas.

 \begin{theorem}
The frugality ratio of the VCG auction for the metric uncapacitated facility location problem is  $3$.
\end{theorem}

\section{Learning-Augmented Frugal Auction}\label{sec:UFL-with-advice}
We now consider the strategic UFL problem within the learning-augmented framework, where the auction has access to potentially inaccurate predictions about the true opening costs of the facilities. Specifically, for each facility \(\facility \in L\), a prediction \(\hat{\open}_\facility\) is provided for its true opening cost \(\open_\facility\). Our objective is to leverage this unreliable information to design truthful auctions that minimize the frugality ratio while remaining robust against inaccurate predictions. 

To achieve this, we propose \mechname, described in Auction~\ref{mech:solution-scaled-frugalUFL}. This auction first identifies the optimal solution based on the predicted opening costs $\hat{\mathbf{\open}}$, denoted by \(\hat{\OPT}\). It then computes modified  opening costs $\open'_\ell(S)$ that scale up $\open_\ell$ for $S = \hat{\OPT}$ and facilities $\ell \in \hat{\OPT}$  with true opening cost that exceeds their predicted opening cost. Finally, the auction selects the solution $S$ that minimizes the scaled cost function $d(U,S) + \sum_{\ell \in S} \hat{\open}_\ell(S)$. 
\begin{algorithm}[h]
\caption{\mechname}
\label{mech:solution-scaled-frugalUFL}
\KwIn{Set of clients $U$, set of facilities $L$, parameter $\epsilon$, predicted opening costs $\hat{\open}_\ell$ for all $\ell \in L$}
\KwOut{Selected facility subset $S^*$ and threshold payments for each facility $\ell \in S^*$}

Compute $\hat{\OPT} \gets \arg\min_{S \subseteq L} \left( d(U,S) + \sum_{\ell \in S} \hat{\open}_\ell \right)$\;

Define the modified cost function for any subset $S \subseteq L$:
\[
\open'_\ell(S) \gets
\begin{cases}
\frac{2}{\epsilon} \cdot \open_\ell, & \text{if } S = \hat{\OPT} \text{ and } \open_\ell > \hat{\open}_\ell \\
\open_\ell, & \text{otherwise}
\end{cases}
\]

Compute:
\[
S^* \gets \arg\min_{S \subseteq L} \left( d(U, S) + \sum_{\ell \in S} \open'_\ell(S) \right)
\]

\Return{$S^*$ and threshold payments for each $\ell \in S^*$}\;
\label{mech:frugalUFL}
\end{algorithm}
We note that, to achieve the claimed results, our auction only needs access to the membership of the predicted optimal solution \(\hat{\OPT}\) and to the opening costs of those facilities. We show that \mechname\ can achieve near-optimal consistency (arbitrarily close to \(1\)) while still maintaining competitive robustness.

\begin{theorem}\label{thm:overall}
    Given any $\epsilon \in [0,2]$ as input, \mechname\ is truthful, $(1+\epsilon)$-consistent, and $\max\left\{5, 3+\frac{2}{\epsilon}\right\}$-robust.
    % Given correct predictions, the payment made by auction~\ref{mech:frugalUFL} is no more than the frugal solution.
\end{theorem}
% \st{typical scaled greedy approaches} 
At first glance, this auction may seem counterintuitive and differs from existing mechanisms that also use predictions to scale costs (e.g.,~\cite{BGT223, XL22,CSV24}). While standard approaches typically scale down predicted costs to prioritize social cost minimization, we scale costs upwards. This reversal is driven by our objective: we aim to minimize the frugality ratio (payments) rather than the social cost (efficiency). Consequently, instead of making the predicted optimal solution more appealing, we make it more sensitive to cost increases.
% The primary rationale for this upward scaling is to de-prioritize facilities for overbidding. 
The primary rationale for this upward scaling is to restrict the potential overbidding the bidder can claim while remaining in the optimal solution, thereby preventing a high threshold price.
By amplifying any possible over-reported opening costs over the predicted ones, the auction gains tighter control over facility payments, which can lead to a lower frugality ratio when the predictions are accurate.

Importantly, the scaling applied by our auction is \emph{solution-dependent} rather than facility-dependent. Specifically, the opening cost scaling is activated only when evaluating the predicted optimal solution \(\hat{\OPT}\) as a complete set. For any other subset, the opening costs of individual facilities remain unscaled. This distinction is crucial for our robustness analysis, as it ensures that the adverse effects of inaccurate predictions are confined to the evaluation of the single predicted optimal solution, leaving alternative solutions unaffected.

We first prove the truthfulness of the auction. We then prove the consistency and robustness.

\subsection{Truthfulness of \mechname}\label{app:latruthful}

We first prove that the auction is truthful. By Myerson's Lemma (Lemma~\ref{lem:myerson}), it suffices to prove the monotonicity of the allocation rule.
\begin{lemma}\label{lem:truthful} For any $\epsilon \leq 2$, the \mechname\ auction is truthful.
% Auction~\ref{mech:solution-scaled-frugalUFL} is truthful for $\epsilon \leq 2$
\end{lemma}
\begin{proof}
We show that the allocation rule of Auction~\ref{mech:solution-scaled-frugalUFL} is monotone, and then invoke Myerson’s lemma to conclude truthfulness.

Fix a facility $\tilde\ell\in L$, and hold constant the reported opening costs $\mathbf{b}_{-\tilde\ell}$ of all other facilities as well as all predictions $\hat{\mathbf{\open}}$.  It suffices to prove that if $\tilde\ell$ belongs to the winning set when it reports cost $a\ge0$, then it still belongs to the winning set when it reports any lower cost $b\le a$.

Recall that the auction uses the modified opening costs
\[
\open'_\ell(S)=
\begin{cases}
\tfrac{2}{\epsilon}\,\open_\ell, & \text{if }S=\hat\OPT\text{ and }\open_\ell>\hat \open_\ell,\\
\open_\ell, & \text{otherwise.}
\end{cases}
\]
We slightly abuse notation and use $\open'_{\tilde\ell}(S,a)$ and $\open'_{\tilde\ell}(S,b)$ to denote the scaled cost when facility $\tilde\ell$ reports $a$ and $b$, respectively. Since lowering $\tilde\ell$’s reported cost can only (weakly) decrease its scaled cost in \emph{every} candidate set, and since the event “$S=\hat\OPT$” depends only on the prediction, not on the reported bid, we have
\[
\open'_{\tilde\ell}(S,b)\;\le\;\open'_{\tilde\ell}(S,a)
\quad\text{for all }S.
\]

Let $S^*$ be the winning set when $\tilde\ell$ bids $a$, and let $S'$ be the winning set when it instead bids $b$, keeping everything else fixed.  If $S'=S^*$ then $\tilde\ell\in S'$ immediately.  Otherwise the only cost that changed is $\open'_{\tilde\ell}$, so in order for the auction to switch to a different optimal set $S'$, it must still include $\tilde\ell$.  Hence in all cases $\tilde\ell$ remains selected, proving monotonicity.  Finally by Myerson’s lemma, a monotone allocation rule induces truthful payments, so the auction is truthful.
\end{proof}

\subsection{Consistency analysis of \mechname}\label{app:consistency}
 We now prove the consistency and robustness for the proposed auction. We first provide the consistency analysis of \mechname. We then provide the robustness guarantee of the auction, which holds regardless of the quality of the predictions.

\begin{lemma}\label{thm:consistency} 
 Given any $\epsilon \in [0,2]$ as input,
 \mechname\ achieves a consistency of $1+\epsilon$. 
    % Given correct predictions, the payment made by Auction~\ref{mech:frugalUFL} is no more than the frugal solution.
\end{lemma}
\begin{proof}Recall that consistency analysis assumes the predictions are correct, i.e., $\hat{\open}_\ell = \open_\ell$ for all facilities $\ell \in L$. Under accurate predictions, the condition $\open_\ell > \hat{\open}_\ell$ is never met for the set $\hat{\OPT}$. Consequently, the mechanism selects the true optimal solution, i.e., $S^* = \hat{\OPT} = \OPT$. To bound the payments, we apply Lemma~\ref{lem:unified-wprime} with the winner set $W = \OPT$, the subset $W' = \OPT$, and the reference set $R = F$. We determine the multipliers as follows:\begin{enumerate}\item \textbf{Winner Multiplier ($\alpha_\ell(W)$):} For any $\ell \in \OPT$, the threshold payment $p_\ell$ represents the maximum bid the agent could report before the solution changes. If an agent reports $b_\ell > \hat{\open}_\ell$, the mechanism scales this cost by $2/\epsilon$. Thus, the effective constraint on the payment is scaled, implying $\alpha_\ell(W) = 2/\epsilon$. In all other cases, the multipliers for Lemma~\ref{lem:unified-wprime} are $\alpha_g(S) = 1$.
\item \textbf{Reference Multiplier ($\alpha_f^{\max}$):} For the reference set $F$, we evaluate the multiplier at the true opening costs. Since predictions are accurate ($\hat{\open}_f = \open_f$), the condition $\open_f > \hat{\open}_f$ is false. Thus, even if $F = \hat{\OPT}$, no scaling applies to the reference costs. Hence, $\alpha_f^{\max} = 1$.\end{enumerate}Substituting these into Lemma~\ref{lem:unified-wprime}:$$\sum_{\ell \in \OPT} \frac{2}{\epsilon}\, p_\ell
\ \le\
\sum_{f \in F} 1 \cdot \open_f \;+\; 2\,d(U,F).$$Using the fact that $\sum_{f \in F} \open_f + 2d(U,F) \le 2 c(F)$, we simplify the inequality:$$\frac{2}{\epsilon} \sum_{\ell \in \OPT} p_\ell \ \le\ 2 c(F) \quad \implies \quad \sum_{\ell \in \OPT} p_\ell \ \le\ \epsilon\, c(F).$$Finally, adding the connection cost of the optimal solution (which satisfies $d(U, \OPT) \le c(F)$), the total cost is bounded by:$$\sum_{\ell \in \OPT} p_\ell + d(U, \OPT) \ \le\ \epsilon\, c(F) + c(F) \ =\ (1+\epsilon)\, c(F). \qedhere$$\end{proof}
We now move on to the robustness analysis. We begin with the simpler case in which the auction still outputs the optimal solution of the instance, i.e., \(S^* = \OPT\). 
\begin{lemma}\label{lem:outputoptrobusyness}
Given any instance $(U, \mathbf{\open}, d)$ and any $\epsilon \leq 2$, let $\OPT$ and $F$ be the optimal solution and frugal solution, respectively. If auction~\ref{mech:solution-scaled-frugalUFL} outputs $S^* = \OPT$, then we have:
\[\sum_{\ell \in S^*} p_\ell + d(U, S^*) \leq \left(1+\frac{2}{\epsilon}\right) \sum_{\ell \in F} \open_\ell + 3\; d(U,F) \]
% Changed index in min from i to \ell for consistency
\end{lemma}
\begin{proof}
By definition of the auction, at most one subset, $\hat{\OPT}$, can have its costs scaled. Therefore, it suffices to consider the relationship between $\hat{\OPT}$ and $\OPT$ (the true optimal solution). We handle the two cases depending on whether the predicted optimal set $\hat{\OPT}$ matches the true optimal set $\OPT$.

\paragraph{Case one: $\hat{\OPT}\neq \OPT$.}
In this case, the true optimum $\OPT$ is not the set subject to scaling, so for the winner set $W = S^* = \OPT$, we have $\alpha_\ell(W) = 1$. However, the frugal set $F$ (or parts of it) might coincide with $\hat{\OPT}$. In the worst case, the alternative solution evaluates $F$ using scaled costs, so $\alpha_f^{\max} \le \frac{2}{\epsilon}$.

We apply Lemma~\ref{lem:unified-wprime} with $W'=\OPT$, $R=F$, $\alpha_\ell(W)=1$, and $\alpha_f^{\max} = \frac{2}{\epsilon}$:
\[
\sum_{\ell\in\OPT} p_\ell\ \le\ \frac{2}{\epsilon}\sum_{f\in F}\open_f\;+\;2\,d(U,F).
\]
To bound the total cost, we add the connection cost $d(U,\OPT)$ to both sides. Using the fact that the connection cost of the optimal solution is bounded by the total cost of the frugal solution ($d(U,\OPT) \le c(\OPT) \le c(F) = \sum_{f\in F}\open_f + d(U,F)$), we obtain:
\begin{align*}
\sum_{\ell\in\OPT} p_\ell + d(U,\OPT) \ &\le\ \frac{2}{\epsilon}\sum_{f\in F}\open_f + 2\,d(U,F) + \left(\sum_{f\in F}\open_f + d(U,F)\right) \\
&=\ \left(1+\frac{2}{\epsilon}\right)\sum_{f\in F}\open_f\;+\;3\,d(U,F).
\end{align*}

\paragraph{Case two: $\hat{\OPT} = \OPT$.}
Since $S^* = \OPT$ and the predicted set matches the optimal set, the consistency analysis from Lemma~\ref{thm:consistency} applies directly. That lemma guarantees a total cost of at most $(1+\epsilon)c(F)$.
Since $\epsilon \le 2$, we have $(1+\epsilon) < (1+\frac{2}{\epsilon})$ and $(1+\epsilon) \le 3$. Thus, the bound from Case one dominates.
\end{proof}
We now address the case where scaling alters the output set, i.e., \(S^* \neq \OPT\). This case is particularly challenging because \(S^*\) may contain facilities belonging to both the frugal solution and the optimal solution, making it difficult to bound the payment. To tackle this, we need a different accounting argument than the ones used for the VCG and consistency analyses. Specifically, we will analyze payments for facilities in the output solution $S^*$ differently, based on whether they are part of the frugal solution $F$.
We begin by establishing a bound for payments to non-frugal facilities in \(S^*\) (i.e., those in \(S^* \setminus F\)) with the following lemma.

\begin{lemma}\label{lem:nonfrualbound} % Consider fixing label typo to nonfrugalbound?
Given any instance $(U, \mathbf{\open},d)$, and any $\epsilon \leq 2$, let $\OPT$ and $F$ be the optimal solution and frugal solution, respectively. Let $S^* \neq \OPT$ be the output of Auction~\ref{mech:solution-scaled-frugalUFL}, and let $S = S^* \setminus F$. Then:
% Changed p_\ell to p_\ell to match summation index
\[\sum_{\ell \in S} p_\ell  \leq  \sum_{\ell \in F} \open_\ell + 2 d(U,F).\]
\end{lemma}
\begin{proof}
Consider the case where $S^* \neq \OPT$. As noted in the analysis of the mechanism, this outcome implies that the predicted optimal set was the true optimal set ($\hat{\OPT} = \OPT$), and its costs were scaled up, causing the mechanism to select an unscaled alternative $S^*$. Consequently, the winner set $S^*$ is not subject to scaling, so $\alpha_\ell(S^*) = 1$ for all $\ell \in S^*$. Furthermore, the frugal solution $F$ is distinct from $\OPT$ (and thus distinct from $\hat{\OPT}$), so facilities in $F$ are also unscaled in any alternative solution, implying $\alpha_f^{\max} = 1$.

We apply Lemma~\ref{lem:unified-wprime} with the winner set $W = S^*$, the subset $W' = S = S^* \setminus F$, and the reference set $R = F$. Substituting the parameters $\alpha_\ell(S^*) = 1$ and $\alpha_f^{\max} = 1$:
\[
\sum_{\ell \in S} 1 \cdot p_\ell \;\le\; \sum_{f \in F} 1 \cdot \open_f \;+\; 2\,d(U,F). \qedhere
\]
\end{proof}
Then, we bound the payment to the output solution that is also part of the frugal solution. 
\begin{lemma}\label{lem:frugalpartbound}
    Given any instance $(U, \mathbf{\open}, d)$, and any $\epsilon \leq 2$, let $\OPT$ and $F$ be the optimal solution and frugal solution, respectively. Let $S^* \neq \OPT$ be the output of Auction~\ref{mech:solution-scaled-frugalUFL}, and $S^f = S^* \cap F$. Then:
    % Changed sum index i to f (as element of S^f). Changed o_\ell to \open_\ell.
    \[\sum_{\ell \in S^f} p_\ell \leq \max\left\{2, \frac{2}{\epsilon}\right\}\cdot \left[ \sum_{\ell \in F} \open_\ell + d(U,F)\right] .\]
\end{lemma}
% \begin{proof}
% % [Proof of Lemma~4.6 (replacement via Lemma~\ref{lem:unified-wprime})]
% From Lemma~\ref{lem:unified-wprime} with $W'=S_f=S^*\cap F$, $R=\OPT$, and
% $\alpha_\ell(W)= 1$, we get
% \[
% \sum_{\ell\in S_f} p_\ell\ \le\ \sum_{g\in\OPT}o_g\;+\;2\,d(U,\OPT)\quad\text{or}\quad
% \sum_{\ell\in S_f} p_\ell\ \le\ \frac{2}{\varepsilon}\sum_{g\in\OPT}o_g\;+\;2\,d(U,\OPT).
% \]
% Using $\sum_{g\in\OPT}o_g\le c(\OPT)\le c(F)$ and $d(U,\OPT)\le c(F)$ gives
% $\sum_{\ell\in S_f}p_\ell \le (\beta+2)c(F)$ with $\beta\in\{1,2/\varepsilon\}$.
% Finally, for facilities in $S_f\cap \OPT$ we may reroute users directly to $\ell\in\OPT$ (the
% “direct routing” case in the draft), improving the $2\,d(U,\OPT)$ term to $d(U,\OPT)$; combining
% the two subcases yields
% \[
% \sum_{\ell\in S_f}p_\ell \ \le\ \max\Bigl\{2,\ \frac{2}{\varepsilon}\Bigr\}\,c(F),
% \]
% as stated. \qedhere
% \end{proof}
\begin{proof}
We apply Lemma~\ref{lem:unified-wprime} with the winner set $W = S^*$, the subset $W'=S^f$, and the reference set $R=\OPT$.For the left-hand side, facilities in $S^f$ are part of the output $S^*$; since the mechanism only scales the specific set $\hat{\OPT}$ and we are in a case where we compare against the true $\OPT$ (which may be scaled in the alternative), the multiplier for the winners is standard: $\alpha_\ell(W) = 1$.For the right-hand side, the reference set is $\OPT$. In the worst case (if $\OPT = \hat{\OPT}$ and costs are under-predicted), the mechanism scales the opening costs of $\OPT$ by $2/\epsilon$. Thus, $\alpha_f^{\max} = 2/\epsilon$. If no scaling triggers, $\alpha_f^{\max} = 1$.Substituting these into Lemma~\ref{lem:unified-wprime}:
$$\sum_{\ell \in S^f} p_\ell \;\le\; \frac{2}{\epsilon} \sum_{j \in \OPT} \open_j \;+\; 2\,d(U, \OPT).$$Since $c(\OPT) = \sum_{j \in \OPT} \open_j + d(U, \OPT)$ and $c(\OPT) \le c(F)$:\begin{itemize}\item If $\frac{2}{\epsilon} \ge 2$ (i.e., $\epsilon \le 1$), the RHS is bounded by $\frac{2}{\epsilon} c(\OPT) \le \frac{2}{\epsilon} \left[ \sum_{\ell \in F} \open_\ell + d(U,F)\right]$.\item If $\frac{2}{\epsilon} < 2$ (i.e., $\epsilon > 1$), the RHS is bounded by $1 \cdot \sum \open_j + 2d(U, \OPT) \le 2 c(\OPT) \le 2 \left[ \sum_{\ell \in F} \open_\ell + d(U,F)\right]$.\end{itemize}combining these cases yields the stated bound.\end{proof}

% The proof is deferred to Appendix~\ref{sec:frugalpartboundproof}. 
By combining Lemma~\ref{lem:outputoptrobusyness}, Lemma~\ref{lem:nonfrualbound}, and Lemma~\ref{lem:frugalpartbound}, we get the robustness of Auction \mechname. 
% The proof is deferred to Appendix~\ref{app:robustness}. 
\begin{lemma}\label{lem:robustness}
   Given any $\epsilon \in [0,2]$ as input, \mechname\ achieves a robustness of $\max\left\{5, 3+\frac{2}{\epsilon}\right\}$.
\end{lemma}
\begin{proof}
    Consider any instance $(U, \mathbf{\open}, d)$ and any $\epsilon \leq 2$. Let $\OPT$ and $F$ be the optimal solution and frugal solution of the given instance. Let $S^*$ be the output of Auction~\ref{mech:solution-scaled-frugalUFL}. 

\paragraph{Case one: $S^* = \OPT$.} If $S^* = \OPT$, by Lemma~\ref{lem:outputoptrobusyness} we have:
\[\sum_{\ell \in S^*} p_\ell + d(U, S^*) \leq \left(\frac{2}{\epsilon}+1\right) \sum_{\ell \in F} \open_\ell + 3 d(U,F). \]
% Corrected min_{i \in F}d(u,i) - removed space

\paragraph{Case two: $S^* \neq \OPT$.} We now consider the case where $S^* \neq \OPT$. First, note that the \emph{connection cost} of $S^*$ is weakly less than or equal to the total cost of any solution $K$, including the frugal solution $F$. We have:
% Corrected: "better than" -> "less than or equal to", "include" -> "including", added comma after "First"
\begin{align}\label{eq:connectioncost}
    d(U, S^*)  \leq \sum_{\ell \in F} \open'_\ell(F) + d(U,F) = \sum_{\ell \in F} \open_\ell + d(U,F),
\end{align}
% Corrected index K to U
where the equality is due to the fact that $\OPT$ is the solution that is scaled, by the fact that $S^* \neq \OPT$.

Let $S =  S^* \setminus F$ and $S^f = S^* \cap F$. Putting together Lemma~\ref{lem:nonfrualbound} and Lemma~\ref{lem:frugalpartbound}, we get:
% Added comma after S^f = S^* \cap F
{\allowdisplaybreaks
\begin{align*}
    \sum_{\ell \in S^*} p_\ell + d(U, S^*) &\leq \sum_{\ell \in S} p_\ell + \sum_{\ell \in S^f} p_\ell + d(U, S^*)\\
    & \leq \sum_{\ell \in F} \open_\ell + 2 d(U,F) + \sum_{\ell \in S^f} p_\ell  + d(U, S^*) \tag{by Lem.~\ref{lem:nonfrualbound}}\\
    & \leq \max\left\{3, \frac{2}{\epsilon}+1\right\}\sum_{\ell \in F} \open_\ell + \max\left\{4, \frac{2}{\epsilon}+2\right\}d(U,F) +  d(U, S^*)  \tag{by Lem.~\ref{lem:frugalpartbound}}\\
    & \leq \max\left\{4, \frac{2}{\epsilon}+2\right\}\sum_{\ell \in F} \open_\ell + \max\left\{5, \frac{2}{\epsilon}+3\right\}d(U,F) \tag{by \eqref{eq:connectioncost}}
    \end{align*}}
Taking the worst case of the two scenarios we get that 
\[ \sum_{\ell \in S^*} p_\ell + d(U, S^*)  \leq \max\left\{5, \;\frac{2}{\epsilon} + 3\right\}\left[\sum_{i \in F} \open_i + d(U,F). \right] \qedhere\] 
% Final formula kept as is, per instructions.
\end{proof}

We are now ready to prove the robustness of Auction \mechname.
% \begin{theorem}\label{thm:robustness}
% Auction \mechname\ achieves robustness of $\max(5, 3+\frac{2}{\epsilon})$.
% \end{theorem}
\begin{proof}[Proof of Theorem~\ref{thm:overall}]
Auction \mechname \ is truthful by Lemma~\ref{lem:truthful}, $(1+\epsilon)$-consistent by Lemma~\ref{thm:consistency}, and $\max\left\{5, \;\frac{2}{\epsilon} + 3\right\}$-robust by Lemma~\ref{lem:robustness}.
\end{proof}

\section{The Error-Tolerant Scaled VCG}\label{sec:error}

Auction~\mechname\ achieves $(1+\epsilon)$-consistency in its frugality ratio when provided with perfect predictions, while maintaining a $\max\left\{5, 3+\frac{2}{\epsilon}\right\}$-robustness guarantee in adversarial scenarios. However, these bounds capture only the extremes of prediction quality. In this section, we extend the auction to achieve an improved frugality ratio for approximately accurate predictions, thereby offering a more granular performance characterization.

We define the prediction error $\eta \ge 1$ as the maximum multiplicative deviation between the predicted and actual opening costs:
% We denote the prediction error as \(\eta \ge 1\), where \(\eta\) is defined as the largest ratio between the predicted opening cost and the actual opening cost, i.e.,
\[
\eta = \max_{\ell \in L} \max\left\{\frac{\hat{o}_\ell}{o_\ell}, \frac{o_\ell}{\hat{o}_\ell}\right\}.
\]
% The auction, \mecherr, formally defined in Auction~\ref{mech:solution‑scaled‑frugalUFL}, extends \mechname\ by incorporating an error-tolerance parameter \(\lambda \ge 0\). Recall that the original auction selects the predicted-optimal facility set \(\hat\OPT\) and, whenever a facility’s true opening cost exceeds its prediction, inflates that cost by a factor of \(2/\epsilon\); otherwise, it leaves opening costs unchanged. 

The proposed auction, \mecherr\ (Algorithm~\ref{mech:solution‑scaled‑frugalUFL}), extends \mechname\ by incorporating an error-tolerance parameter \(\lambda \ge 1\). The mechanism modifies the cost function $c_\lambda(S)$ based on the accuracy of the reported costs relative to the predictions.

\begin{algorithm}[ht]
\caption{\mecherr}
\label{mech:solution‑scaled‑frugalUFL}
\KwIn{Clients \(U\), facilities \(L\), predicted costs \(\hat{o}_\ell\) for all \(\ell\in L\), parameter $\epsilon \leq 2$, error‑tolerance \(\lambda\ge0\)}
\KwOut{Chosen facilities \(S^*\) and threshold payments for each \(\ell\in S^*\)}

%---------------------------------------------------
% 1. Compute predicted-optimal solution
%---------------------------------------------------
\(\;\;\hat{\OPT}\;\gets\;\arg\min_{S\subseteq L}\Bigl(d(U,S)\;+\;\sum_{\ell\in S}\hat{o}_\ell\Bigr)\)\;

%---------------------------------------------------
% 2. Check if all true costs lie within factor λ
%---------------------------------------------------
\uIf{$ \forall \ell \in \hat{\OPT}$, $\open_\ell \leq \lambda \hat{\open}_{\ell}$}{
  % scaled‐whole‐solution case
  \[
    \text{Define}\quad c_\lambda(S)\;\;=\;
    \begin{cases}
      \displaystyle\frac{1}{\lambda^2}\Bigl(d(U,S)\;+\;\sum_{\ell\in S}o_\ell\Bigr),
      & S=\hat{\OPT},\\[1em]
      \displaystyle d(U,S)\;+\;\sum_{\ell\in S}o_\ell,
      & \text{otherwise.}
    \end{cases}
  \]
}
\Else{
  % per‐facility inflation case (for large true cost)
  \[
    \text{ Define for each $\ell$ and $S$}\quad o'_\ell(S)\;=\;
    \begin{cases}
      \tfrac{2}{\epsilon}\,o_\ell,
      & S=\hat{\OPT}\text{ and }o_\ell> \lambda \hat{o}_\ell,\\
      o_\ell, & \text{otherwise,}
    \end{cases}
  \]

  \[
    \text{and let} \quad c_\lambda(S)\;=\;d(U,S)\;+\;\sum_{\ell\in S}o'_\ell(S).
  \]
}

%---------------------------------------------------
% 3. Select and pay
%---------------------------------------------------
\[
  S^*\;\gets\;\arg\min_{S\subseteq L}c_\lambda(S),
\]
\Return{\(S^*\) and threshold payments for each \(\ell\in S^*\)}\;
\end{algorithm}

\mecherr\ is designed to output $\hat{\OPT}$ not only under exact predictions but whenever predictions are accurate within a factor of \(\lambda\). To achieve this, we employ a two-stage verification. If every facility in \(\hat\OPT\) reports a true cost $o_\ell \leq \lambda \hat{o}_\ell$, the mechanism applies a uniform down-scaling of $1/\lambda^2$ to the total cost of $\hat{\OPT}$ (including connection costs). This aggressive scaling ensures that $\hat{\OPT}$ remains the minimizer of $c_\lambda(S)$ even if errors up to $\eta \le \lambda$ exist. Conversely, if any facility in $\hat{\OPT}$ deviates beyond $\lambda$, the mechanism reverts to the specific penalty logic of \mechname, inflating the costs of the specific outliers by $2/\epsilon$. Theoretically, \mecherr\ achieves an approximation guarantee of $\eta(1+\lambda) + 2\epsilon$ when \(\eta \le \lambda\), while maintaining a robustness bound of \(\max\{2\lambda^4 + 3\lambda^2, 3 + 2/\epsilon\}\) for arbitrary errors.

% \mecherr\ is designed to preserve this behavior not only under exact predictions but also when predictions are accurate up to a factor of \(\lambda\), with one key enhancement. To ensure that \(\hat\OPT\) remains the chosen set under low prediction error, we replace the overscaling threshold \(\hat{o}_\ell\) with \(\lambda\,\hat{o}_\ell\). In addition, if every facility in \(\hat\OPT\) reports a true cost below this new threshold, we apply a uniform down-scaling to the entire solution cost (both opening and connection costs). This extra adjustment guarantees that, when the prediction error is at most \(\lambda\), the auction outputs \(\hat\OPT\). Aside from these two modifications, \mecherr\ follows \mechname\ exactly. It achieves an approximation guarantee of
% $
% \eta\,(1+\lambda) + 2\epsilon
% $
% when \(\eta \le \lambda\), and maintains a frugality bound of \(\max\{2\lambda^4 + 3\lambda^2, 3 + 2/\epsilon\}\) for arbitrary error \(\eta\).
\begin{theorem}\label{thm:combined-frugality}
Auction \mecherr\ is truthful, and given parameters $\epsilon\in (0, 2]$ and $\lambda>1$, it achieves the following frugality ratio, where $\eta$ is the error of the prediction:
\[
\begin{cases}
\displaystyle
\eta(1+\lambda)+2\epsilon,
&\text{if }\eta \le \lambda,\\[1em]
\displaystyle
\max \ \!\Bigl\{\,2\lambda^4+3\lambda^2,\;3+\tfrac{2}{\epsilon}\Bigr\},
&\text{if }\eta > \lambda.
\end{cases}
\]
\end{theorem}
\subsection{Truthfulness of \mecherr}
We first prove the truthfulness of Auction~\mecherr. The proof is very similar to Theorem~\ref{lem:truthful}.
\begin{lemma}\label{lem:truthful1}
    For any $\epsilon \leq 2$, the \mecherr\ auction is Truthful.
\end{lemma}
\begin{proof}
We show that the allocation rule of Auction~\ref{mech:solution‑scaled‑frugalUFL} is monotone.

Fix a facility $\tilde\ell\in L$, and $\mathbf{o}_{-\tilde\ell}$ of all other facilities as well as all predictions $\hat{\mathbf{o}}$.  Let $\tilde\ell$’s reported cost be $a\ge0$, and suppose that when it bids $a$ the auction selects $S^*_a\ni\tilde\ell$.  We must show that if it instead reports any lower cost $b\le a$, it remains selected.

Denote by $c_\lambda(S\,;\,x)$ the total cost of set $S$ under report $x\in\{a,b\}$, according to Auction~\ref{mech:solution‑scaled‑frugalUFL}.  Note that in either branch (the “whole‑solution downscaling” or the “per‑facility inflation”), lowering $\tilde\ell$’s cost from $a$ to $b$ \emph{weakly decreases} its contribution to $c_\lambda(S)$ for \emph{every} $S$.  We consider two cases:

\smallskip
\noindent\textbf{Case 1: Whole‑solution downscaling applies at report $a$.}  That is, 
\[
\forall\,\ell'\in\hat\OPT,\;o_{\ell'}\le\lambda\,\hat o_{\ell'}\quad\text{under }x=a.
\]
Since $b\le a$, the same test holds at $x=b$, so the auction remains in the downscaling branch.  In that branch
\[
c_\lambda(S\,;\,x)
=\begin{cases}
\frac1{\lambda^2}\Bigl(d(U,S)+\sum_{\ell\in S}o_\ell\Bigr),&S=\hat\OPT,\\
d(U,S)+\sum_{\ell\in S}o_\ell,&\text{otherwise},
\end{cases}
\]
and reducing $\tilde\ell$’s cost from $a$ to $b$ multiplies its term by at most~$b/a\le1$ in every candidate set.  Hence 
\[
c_\lambda(S\,;\,b)\;\le\;c_\lambda(S\,;\,a)
\quad\text{for all }S,
\]
Since solution that got a reduced cost must contain $\tilde\ell$,  any new minimizer $S^*_b$ must still contain~$\tilde\ell$.

\smallskip
\noindent\textbf{Case 2: Per‑facility inflation applies at report $a$.}  Then there is some $\ell'\in\hat\OPT$ with $o_{\ell'}> \lambda\hat o_{\ell'}$ under $x=a$.  Two subcases arise:

- \emph{If $\tilde\ell\notin\hat\OPT$.}  Lowering its bid does not affect the branch test, so we stay in the inflation branch.  But in that branch each facility’s scaled cost (whether inflated by $2/\epsilon$ or not) is a nondecreasing function of its report, so the argument from the original auction applies verbatim to show monotonicity.

- \emph{If $\tilde\ell\in\hat\OPT$.}  Then $a>\lambda\hat o_{\tilde\ell}$ but $b\le\lambda\hat o_{\tilde\ell}$, so at $x=b$ the auction flips into the downscaling branch.  In that branch it will choose $\hat\OPT$ (since we uniformly downscale its cost), and $\hat\OPT\ni\tilde\ell$.  Thus $\tilde\ell$ remains selected.

\smallskip
In all cases, lowering $\tilde\ell$’s bid can only weakly decrease its cost in every candidate set including it, and any switch of the minimizer continues to include~$\tilde\ell$.  Therefore the allocation rule is monotone, and by Myerson’s lemma the resulting payments make the auction truthful.
\end{proof}

\subsection{Performance analysis for \mecherr}
We now provide the performance as a function of the error $\eta$ and the error tolerance parameter $\lambda$.

\subsubsection{Performance when $\eta \leq \lambda$}
\label{sec:perfone}

We first consider the case where $\eta \leq \lambda$, we start by showing the claim that $\hat{\OPT}$ is always outputted when $\eta \leq \lambda$. The result of this case is summarized in Lemma~\ref{thm:frugality‑bound_withinerror}. We first show that under the condition $\eta \leq \lambda$, the prediction optimal is always returned.
% After computing
% \[
% \hat\OPT \;=\;\arg\min_{S\subseteq L}\Bigl(d(U,S)\;+\;\sum_{\ell\in S}\hat{o}_\ell\Bigr),
% \]
% the auction checks whether
% \[
% \max_{\ell\in\hat\OPT}\frac{o_\ell}{\hat{o}_\ell}\;\le\;\lambda.
% \]
% \begin{itemize}
%   \item If this holds, it “solution‑level” scales the entire cost of \(\hat\OPT\)—both connection and opening costs—by a factor of \(1/\lambda\).
%   \item Otherwise, it defaults to the per‑facility inflation rule: for each \(\ell\in\hat\OPT\) with \(o_\ell>\hat{o}_\ell\), it replaces \(o_\ell\) by \(\tfrac{2}{\epsilon}o_\ell\), leaving all other costs intact.
% \end{itemize}
% The remainder of the auction is identical to the base version: it chooses
% \[
% S^* \;=\;\arg\min_{S\subseteq L}c_\lambda(S),
% \]
% where \(c_\lambda\) is the modified cost function above, and then issues the usual VCG‑style threshold payments to the opened facilities.

\begin{lemma}\label{lem:outputerr}
If \(\eta \le \lambda\), then Auction~\ref{mech:solution‑scaled‑frugalUFL} always selects \(\hat\OPT\); that is, \(S^*=\hat\OPT\).
\end{lemma}

We now provide an upper bound of the connection cost of the output solution.

\begin{proof}
Let
\[
\hat{c}(S)\;=\;d(U,S)\;+\;\sum_{\ell\in S}\hat o_\ell,
% \qquad
% c(S)\;=\;d(U,S)\;+\;\sum_{\ell\in S}o_\ell,
\]
be the predicted total cost. First note that since  \(\eta\le\lambda\), we have $\open_\ell \leq \lambda \hat{\open_\ell}$ for all $\ell$ the auction will set
\[
c_\lambda(S)\;=\;
\begin{cases}
\displaystyle\frac{1}{\lambda^2}\,c(\hat\OPT),&S=\hat\OPT,\\
c(S),&S\neq\hat\OPT.
\end{cases}
\]
Since \(\hat\OPT\) minimizes the \emph{predicted} cost,
\begin{align}\label{eq:minimizepredicted}
    \hat{c}(\hat\OPT)\;\le\;\hat{c}(S)\quad\forall\,S\subseteq L.
\end{align}

By \(\eta\le\lambda\) we have \(o_\ell\le\lambda\,\hat o_\ell\) for all \(\ell\in\hat\OPT\), so
\[
c(\hat\OPT)
=\sum_{u}d(u,\hat\OPT)\;+\;\sum_{\ell\in\hat\OPT}o_\ell
\;\le\;
\sum_{u}d(u,\hat\OPT)\;+\;\lambda\sum_{\ell\in\hat\OPT}\hat o_\ell
\leq \lambda\,\hat{c}(\hat\OPT),
\]
% i.e.\ 
% \(\displaystyle \frac{1}{\lambda}c(\hat\OPT)\le \hat{c}(\hat\OPT)\le \hat{c}(S)\).
On the other hand, since \(\hat o_\ell\le\eta\,o_\ell\) for all \(\ell\), we get
\[
\hat{c}(S)
=\sum_{u}d(u,S)\;+\;\sum_{\ell\in S}\hat o_\ell
\;\le\;
\sum_{u}d(u,S)\;+\;\eta\sum_{\ell\in S}o_\ell
=\eta\,c(S),
\]
Combining the above inequality with \eqref{eq:minimizepredicted} we have,
\[
c(\hat\OPT)\;\le\; \lambda \hat{c}(S)\;\le\;\lambda \eta\,c(S)
\;\implies\;
\frac{1}{\lambda^2}c(\hat\OPT)\;\le\;c(S) \;\implies\; c_\lambda(\hat{\OPT}) \leq c_\lambda(S),\quad\forall\,S\neq\hat\OPT.
\]
Since the auction output the set that minimizr $c_\lambda$, we therefore have \(S^*=\hat\OPT\).
\end{proof}

\begin{observation}\label{obs:etabound}
    For any set $S \neq \hat{\OPT}$, let $\hat{c}(S)\;=\;d(U,S)\;+\;\sum_{\ell\in S}\hat o_\ell$, we have
    \[\hat{c}(S) \leq \eta c(S).\]
\end{observation}
\begin{proof}
    \(\eta \le \lambda\) implies \(\hat o_\ell \le \lambda\,o_\ell\) for every \(\ell\), so
\[
\hat{c}(S)
=
d(U,S)
\;+\;\sum_{\ell\in S}\hat o_\ell
\;\le\;
d(U,S)
\;+\;\eta\sum_{\ell\in S}o_\ell
\;=\;
\eta \Bigl(d(U,S)+\sum_{\ell\in S}o_\ell\Bigr).
\qedhere\]
\end{proof}

We now provide an upper bound of the connection cost of the output solution.

\begin{lemma}\label{lem:connection}
If \(\eta \le \lambda\), then for any \(S\subseteq L\),
\[
d\left(U, \hat{\OPT}\right)
\;\le\;
\eta \;\Bigl(d(U,S)\;+\;\sum_{\ell\in S}o_\ell\Bigr).
\]
\end{lemma}

\begin{proof}
% Define the \emph{predicted cost}
% \[
% \hat{c}(T)\;=\;d(U,T)\;+\;\sum_{\ell\in T}\hat o_\ell.
% \]
% Since \(\hat\OPT\) minimizes \(\hat{c}\), we have
% \[
% \hat{c}(\hat\OPT)\;\le\;\hat{c}(S).
% \]
% First, \(\eta \le \lambda\) implies \(\hat o_\ell \le \lambda\,o_\ell\) for every \(\ell\), so
% \[
% \hat{c}(S)
% =
% d(U,S)
% \;+\;\sum_{\ell\in S}\hat o_\ell
% \;\le\;
% d(U,S)
% \;+\;\lambda\sum_{\ell\in S}o_\ell
% \;=\;
% \lambda\Bigl(d(U,S)+\sum_{\ell\in S}o_\ell\Bigr).
% \]
Since \(d\left(U, \hat{\OPT}\right)\le \hat{c}(\hat\OPT)\), combining with the fact that \(\hat\OPT\) minimizes \(\hat{c}\), we have gives
\[
d\left(U, \hat{\OPT}\right)
\;\le\;
\hat{c}(\hat\OPT)
\;\le\;
\hat{c}(S)
\;\le\;
\eta\Bigl(d(U,S)+\sum_{\ell\in S}o_\ell\Bigr),
\]
where the last inequality is by Observation~\ref{obs:etabound}.
\end{proof}

We now move on to upper bound the payment to the facilities in the output solution, i.e., $\hat{\OPT}$. In particular, we will partition $\hat{\OPT}$ based on the threshold they applied, and whether it is a member of $F$, the true frugal solution. We first analyze the low-threshold facilities.

\begin{lemma}\label{lem:lowpayment}
Under the condition \(\eta \le \lambda\), let
$
S^1 \;=\;\bigl\{\ell\in\hat\OPT : p_\ell \le \lambda\,\hat o_\ell\bigr\}.
$
Then
\[
\sum_{\ell\in S^1} p_\ell \;\le\;\lambda\,\eta\,c(F),
\]
 where \(F\) is the frugal solution.
% and \(c(F)=d(U,F)+\sum_{\ell\in F}o_\ell\).
\end{lemma}

\begin{proof}
Since \(p_\ell\le\lambda\,\hat o_\ell\) for every \(\ell\in S^1\),
\[
\sum_{\ell\in S^1}p_\ell
\;\le\;
\lambda\sum_{\ell\in S^1}\hat o_\ell
\;\le\;
\lambda\sum_{\ell\in\hat\OPT}\hat o_\ell
\;\le\;
\lambda\,\hat{c}(\hat\OPT)
\;\le\;
\lambda\,\hat{c}(F),
\;\le\;
\lambda\ \eta \;c(F),
\]
where the last inequality is by Observation~\ref{obs:etabound}.
% Finally, since \(\hat o_\ell\le\eta\,o_\ell\) for all \(\ell\), \xnote{the line below appeared many times, so consider merging}
% \[
% \hat{c}(F)
% \;=\;\sum_{u}d(u,F)+\sum_{\ell\in F}\hat o_\ell
% \;\le\;
% \sum_{u}d(u,F)+\eta\sum_{\ell\in F}o_\ell
% \;=\;\eta\,c(F).
% \]
% Combining gives \(\sum_{\ell\in S^1}p_\ell\le\lambda\,\eta\,c(F)\).
\end{proof}

\begin{lemma}\label{lem:nonfrugalhigh}Assume $\eta\le\lambda$, and let$$S^2 \;=\;\bigl\{\ell\in\hat\OPT : p_\ell>\lambda\,\hat o_\ell,\;\ell\notin F\bigr\}.$$Then$$\sum_{\ell\in S^2}p_\ell \;\le\;\epsilon\,c(F),$$where $F$ is the frugal solution.\end{lemma}

\begin{proof}
We apply Lemma~\ref{lem:unified-wprime} with the winner set $W = \hat{\OPT}$, the subset $W'=S^2$, and the reference set $R=F$. 
For every $\ell \in S^2$, the condition $p_\ell > \lambda \hat{o}_\ell$ implies that the mechanism applies the inflation scaling $\alpha_\ell(W) = 2/\epsilon$.
For the facilities in $F$, the cost multiplier in the alternative solution is unscaled, so $\alpha_{f}^{\max} = 1$.
Substituting these parameters into the lemma yields:$$\sum_{\ell\in S^2} \frac{2}{\epsilon}\, p_\ell
\ \le\
\sum_{f\in F}o_f\;+\;2\,d(U,F).$$Noting that $\sum_{f\in F}o_f + 2d(U,F) \le 2(\sum_{f\in F}o_f + d(U,F)) = 2c(F)$, we conclude:$$\sum_{\ell \in S^2} p_\ell \ \le\ \frac{\epsilon}{2} \cdot 2c(F) \ =\ \epsilon\,c(F). \qedhere$$\end{proof}

\begin{lemma}\label{lem:frugalhigh}
Under the condition \(\eta \le \lambda\), let
\[
S^3 \;=\;\bigl\{\ell\in\hat\OPT : p_\ell > \lambda\,\hat o_\ell,\;\ell\in F\bigr\}.
\]
Then
\[
\sum_{\ell\in S^3} p_\ell \;\le\;\epsilon\, \cdot c(\OPT).
\]
% where \(c(\OPT)=d(U,\OPT)+\sum_{\ell\in\OPT}o_\ell\).
\end{lemma}
\begin{proof}
We invoke Lemma~\ref{lem:unified-wprime} with the following parameters: Since $\eta \le \lambda$, the auction selects $W = \hat{\OPT}$. Let $W' = S^3 = \{\ell \in \hat{\OPT} \cap F : p_\ell > \lambda \hat{o}_\ell\}$.
Define the Reference Set ($R$): Let $R = \OPT$ (the true optimal solution).

% 1. Determine Winner Multipliers ($\alpha_\ell(W)$):
For any facility $\ell \in S^3$, the threshold payment $p_\ell$ satisfies $p_\ell > \lambda \hat{o}_\ell$. According to the ERRORTOLERANT mechanism, when a facility in $\hat{\OPT}$ reports a cost exceeding $\lambda \hat{o}_\ell$, the mechanism applies the ``Per-facility inflation'' rule. Thus, the cost multiplier for these facilities is:$$\alpha_\ell(W) \;=\; \frac{2}{\epsilon}.$$
% 2. Determine Reference Multipliers ($\alpha_f^{\max}$):

We now bound the multiplier for any facility $f \in \OPT$. The condition $\eta \le \lambda$ implies that for all facilities, the true cost is close to the prediction ($o_f \le \lambda \hat{o}_f$). Consequently, facilities in $\OPT$ do not trigger the high-cost inflation condition (which requires bid $> \lambda \hat{o}_f$). Their multiplier in the mechanism is at most 1 (either 1 or $1/\lambda^2$). Thus:$$\alpha_f^{\max} \;\le\; 1.$$
Apply Lemma~\ref{lem:unified-wprime} and Substituting these values into the unified bound:
\begin{align*}\sum_{\ell \in S^3} \frac{2}{\epsilon} p_\ell\le \sum_{f \in \OPT} (1) \cdot o_f + 2d(U, \OPT) \;\le 2\left(\sum_{f \in \OPT} o_f + d(U, \OPT)\right) \;= 2c(\OPT).\end{align*}Multiplying both sides by $\frac{\epsilon}{2}$, we obtain:$$\sum_{\ell \in S^3} p_\ell \;\le\; \epsilon\,c(\OPT). \qedhere$$\end{proof}
% \begin{proof}
% For \(\ell\in S^3\), \(\ell\in F\) and \(p_\ell>\lambda\,\hat o_\ell\).  An argument analogous to Lemma 4.5 shows that inflating the opening cost of any facility in the frugal solution beyond its predicted cost by up to a \(2/\epsilon\) factor bounds its threshold payment by \(\epsilon\) times the cost of the true optimum.  Hence \(\sum_{\ell\in S^3}p_\ell\le\epsilon\,c(\OPT)\).
% \end{proof}

\begin{lemma}
\label{thm:frugality‑bound_withinerror}
Assume \(\eta \le \lambda\).  Let \(F\) be the frugal solution, 
% with cost
% \[
% c(F)\;=\;d(U,F)\;+\;\sum_{\ell\in F}o_\ell,
% \]
and let \(S^*\) and \(\{p_\ell\}_{\ell\in S^*}\) be the output of Auction~\ref{mech:solution‑scaled‑frugalUFL}.  Then
\[
d(U,S^*) \;+\;\sum_{\ell\in S^*}p_\ell
\;\le\;\bigl(\eta(1+\lambda)+2\epsilon\bigr)\, \cdot c(F).
\]
\end{lemma}

\begin{proof}
By Lemma~\ref{lem:outputerr}, under \(\eta\le\lambda\) the auction selects \(S^*=\hat\OPT\).  We partition the payments over \(\hat\OPT\) into three parts \(S^1,S^2,S^3\) as in Lemmas\ref{lem:lowpayment}, Lemma~\ref{lem:nonfrugalhigh} and Lemma~\ref{lem:frugalhigh}.
since \(c(\OPT)\le c(F)\).
Summing connection cost and all payments gives
\begin{align*}
    d(U,S^*) \;+\;\sum_{\ell\in S^*}p_\ell
&=\;d(U,\hat{\OPT})\;+\;\sum_{\ell\in S^1}p_\ell
\;+\;\sum_{\ell\in S^2}p_\ell
\;+\;\sum_{\ell\in S^3}p_\ell\\
&\;\le\;
\eta \;c(F)
\;+\;\sum_{\ell\in S^1}p_\ell
\;+\;\sum_{\ell\in S^2}p_\ell
\;+\;\sum_{\ell\in S^3}p_\ell \tag{by Lem.~\ref{lem:connection}}\\
&\;\le\;
\eta\,c(F)
\;+\;\lambda\eta\,c(F)
\;+\;\sum_{\ell\in S^2}p_\ell
\;+\;\sum_{\ell\in S^3}p_\ell\tag{by Lem.~\ref{lem:lowpayment}}\\
&\;\le\;
\eta\,c(F)
\;+\;\lambda\eta\,c(F)
\;+\;\epsilon\,c(F)
\;+\;\sum_{\ell\in S^3}p_\ell\tag{by Lem.~\ref{lem:nonfrugalhigh}}\\
&\;\le\;
\eta\,c(F)
\;+\;\lambda\eta\,c(F)
\;+\;\epsilon\,c(F)
\;+\;\epsilon\,c(F)\tag{by Lem.~\ref{lem:frugalhigh}}\\
&\;=\;
\bigl(\eta(1+\lambda)+2\epsilon\bigr)\, \cdot c(F). \qedhere
\end{align*}
\end{proof}

% Notes: For the case where $\eta \geq \lambda$ proof sketch:
% \begin{enumerate}
%     \item Scaling, no matter which one, is only applied to one solution, $hat{\OPT}$
%     \item We will break it down to cases applied by the auction. case 1 applied $\frac{1}{\lambda}$ on the whole , case 2 applied per-agent scale up $\frac{2}{\epsilon}$.
%     \item For case 1: if $\hat{\OPT}$ is output, we have: the following few things: $D(U, \hat{\OPT}) \leq \lambda^2 c(F)$; similarly break the the facilities down to the the ones in frugal and not in frugal, we get the bound of the payment to each individually is $(3\lambda^2 -1)$, and together we pay $7\lambda^2 -1$
%     \item for case 1: if $\hat{\OPT}$ is not output, the regular VCG bound should apply, because from the winner's perspective, no scale is applied, and their payment, if anything is lower than VCG since one of the competition set is made better by the scaling.
%     \item For case 2: for case two, the analysis is the same with the robustness analysis, because the analysis is independent of wheen the scale is applied, in the worst case it'll be applied.
%     \item so the final bound we are getting should be $\max(7\lambda^2-1, \max(3+\frac{2}{\epsilon}),5)$
%     \end{enumerate}

\subsubsection{Performance for general error $\eta$}
\label{sec:perftwo}

We now analyze performance for arbitrary $\eta$ by breaking the proof into two dimensions: (a) which scaling rule the auction applies: either whole‐solution downscaling or per‐facility inflation—and (b) which set is ultimately selected. In each of the resulting subcases, we separately bound the connection cost and the total payment using rerouting arguments and the appropriate scaling factor; combining these four analyses yields the unified frugality guarantee stated in Lemma~\ref{lem:robust-frugality}. We start by consider the case where \emph{whole-solution scaling} is applied and the $\hat{\OPT}$ is outputed.

\begin{lemma}\label{lem:connectionoverall}Let $S^*$ be the output of Auction~\ref{mech:solution‑scaled‑frugalUFL}, if $S^* = \hat{\OPT}$,
for any \(S \neq \hat{\OPT}\),
\[
d\left(U, S^*\right)
\;\le\;
\lambda^2 \;c(S).
\]
\end{lemma}

\begin{proof}
By the scaling scehem of the auction we have \(d\left(U, \hat{\OPT}\right)\le \lambda^2 c_\lambda(\hat\OPT)\), combining with the fact that \(S^*\) minimizes \(c_\lambda\), we have gives
\[
d\left(U, S^*\right)
\;\le\;
\lambda^2 \cdot c_\lambda(S^*)
\;\le\;
\lambda^2 \cdot c_\lambda(S)
\;\le\;
\lambda^2 \cdot c(S)
\]
where the last inequality is since $S \neq \hat{\OPT}$.
\end{proof}

\begin{lemma}\label{lem:paymenttoopt}
Let $S^*$ be the output of Auction~\ref{mech:solution‑scaled‑frugalUFL}.  If $S^* = \hat\OPT$ and the whole-solution-scaling is applied, then
\[
\sum_{\ell\in S^*}p_\ell \;\le\;(2\lambda^4+2\lambda^2)\,c(F),
\]
% where $F$ is the frugal solution and 
% \[
% c(F)=d(U,F)\;+\;\sum_{\ell\in F}o_\ell.
% \]
\end{lemma}
\begin{proof}We follow the rerouting argument of Lemma~\ref{lem:nonfrualbound}. Partition $S^* = \hat\OPT$ into$$S \;=\;S^* \setminus F
\quad\text{and}\quad
S^f \;=\;S^*\cap F.$$
Since the auction applies a $1/\lambda^2$ down-scaling to the total cost of $\hat\OPT$ (including connection costs), the threshold payment $p_\ell$ for any $\ell \in S$ satisfies:
\begin{equation}\label{eq:peragenterror1}
\frac1{\lambda^2}\Bigl(p_\ell + \sum_{u \in U_\ell} d(u, \ell)\Bigr)\le\sum_{f\in O_\ell \setminus S^*} o_f + \sum_{u \in U_\ell} \min_{v \in O_\ell} d(u, v).
\end{equation}Summing this inequality over all $\ell \in S$ and applying the standard rerouting upper bound (as formalized in the proof of Lemma~\ref{lem:unified-wprime}) with respect to the reference set $F$:$$\frac1{\lambda^2}\Bigl(\sum_{\ell \in S} p_\ell + d(U, S)\Bigr) \;\le\; \sum_{f \in F} o_f + 2d(U,F) + d(U,S).$$Rearranging to isolate the payments:\begin{align*}\frac{1}{\lambda^2} \sum_{\ell \in S} p_\ell;&\le; \sum_{f \in F} o_f + 2d(U,F) + \left(1 - \frac{1}{\lambda^2}\right) d(U, S) \;&\le; 2c(F) + \left(1 - \frac{1}{\lambda^2}\right) d(U, S).\end{align*}Multiplying by $\lambda^2$:$$\sum_{\ell \in S} p_\ell \;\le\; 2\lambda^2 c(F) + (\lambda^2 - 1) d(U, S).$$From Lemma~\ref{lem:connectionoverall} (connection cost bound), we know that under whole-solution scaling, $d(U, S^*) \le \lambda^2 c(F)$. Since $S \subseteq S^*$, $d(U, S) \le d(U, S^*) \le \lambda^2 c(F)$. Substituting this bound:$$\sum_{\ell \in S} p_\ell \;\le\; 2\lambda^2 c(F) + (\lambda^2 - 1)\lambda^2 c(F) \;=\; (\lambda^4 + \lambda^2) c(F).$$Applying the identical argument to the subset $S^f$ (with reference set $\OPT$ instead of $F$) yields the bound $(\lambda^4 + \lambda^2) c(\OPT)$. Since $c(\OPT) \le c(F)$, we sum the two parts to obtain:$$\sum_{\ell\in S^*}p_\ell
\;=\;\sum_{\ell\in S}p_\ell+\sum_{\ell\in S^f}p_\ell
\;\le\;(2\lambda^4+2\lambda^2)\,c(F). \qedhere$$\end{proof}

\begin{lemma}
\label{lem:robust-frugality}
The Auction~\ref{mech:solution‑scaled‑frugalUFL} achieves a frugality ratio
\[
\max\!\Bigl\{\,2\lambda^4+3\lambda^2 ,\;3 + \tfrac{2}{\epsilon}\Bigr\}.
\]
\end{lemma}
\begin{proof}
The auction uses exactly one of two branches:

\textbf{(i) Whole-solution downscaling:}
\[
C_\lambda(S)=
\begin{cases}
\frac{1}{\lambda^2}\Bigl(d(U,S)+\sum_{\ell\in S}o_\ell\Bigr), & S=\hat\OPT,\\
d(U,S)+\sum_{\ell\in S}o_\ell, & S\neq\hat\OPT.
\end{cases}
\]

\textbf{(ii) Per-facility inflation:}
\[
C_\lambda(S)=d(U,S)+\sum_{\ell\in S}o'_\ell(S),
\]
where \(o'_\ell(S)=\frac{2}{\epsilon}\,o_\ell\) if \(S=\hat\OPT\) and \(o_\ell>\lambda\hat o_\ell\), and \(o'_\ell(S)=o_\ell\) otherwise.

\medskip
\noindent\textbf{Case 1: Whole-solution downscaling applies.}

\noindent\textit{Case 1.1:} The auction outputs \(S^*=\hat\OPT\).
By Lemma~\ref{lem:connectionoverall}, the connection cost satisfies \(d(U,S^*) \le \lambda^2\,c(F)\).
Combined with Lemma~\ref{lem:paymenttoopt}, which gives \(\sum_{\ell\in S^*}p_\ell\le(2\lambda^4+2\lambda^2)\,c(F)\), we obtain
\[
  d(U,S^*) + \sum_{\ell\in S^*}p_\ell
  \;\le\;(2\lambda^4+3\lambda^2)\,c(F).
\]

\noindent\textit{Case 1.2:} The auction outputs \(S^*\neq\hat\OPT\).
In this subcase, \(\hat\OPT\) was downscaled but not chosen, and all other candidate sets remain unscaled. Thus, from the perspective of the winners in $S^*$, the outcome is no worse than the standard VCG result. By Corollary~\ref{cor:vcg_frugality}:
\[
  d(U,S^*) \;+\;\sum_{\ell\in S^*}p_\ell
  \;\le\;3\,c(F).
\]

\medskip
\noindent\textbf{Case 2: Per-facility inflation applies.}
In this branch, the analysis mirrors the robustness proof of Theorem~\ref{thm:overall} (the PREDICTEDLIMITS auction). The exact triggering condition (\(o_\ell>\hat o_\ell\) vs.\ \(o_\ell>\lambda\hat o_\ell\)) does not affect the worst-case bound analysis involving \(o'_\ell(S)\). Thus:
\[
  d(U,S^*) + \sum_{\ell\in S^*}p_\ell
  \;\le\;
  \max\Bigl\{5,\;3+\frac{2}{\epsilon}\Bigr\}\,c(F).
\]

\medskip
\noindent\textbf{Conclusion.}
Taking the maximum over all branches yields
\[
\max\Bigl\{2\lambda^4+3\lambda^2,\;5,\;3+\frac{2}{\epsilon}\Bigr\}\,c(F).
\]
Since \(\lambda \ge 1\), \(2\lambda^4+3\lambda^2 \ge 5\), so the term 5 is redundant. The bound simplifies to \(\max\{2\lambda^4+3\lambda^2,\; 3+\frac{2}{\epsilon}\}\,c(F)\).
\end{proof}

\subsubsection{Main lemma for the performance of \mecherr}
Combining Section~\ref{sec:perfone} and Section~\ref{sec:perftwo} gives  the following performance for auction \mecherr.

\begin{lemma}
Given parameters $\epsilon\in(0,2]$ and $\lambda>1$, auction \mecherr \ achieves the following frugality ratio, where $\eta$ is the prediction error:
\[
\begin{cases}
\eta(1+\lambda)+2\epsilon, & \text{if } \eta \le \lambda, \\
\max\{2\lambda^4+3\lambda^2,\; 3+\frac{2}{\epsilon}\}, & \text{if } \eta > \lambda.
\end{cases}
\]
\end{lemma}

\begin{proof}
If $\eta \leq \lambda$, Lemma~\ref{thm:frugality‑bound_withinerror} establishes the bound $\eta(1+\lambda)+2\epsilon$.
For arbitrary $\eta$ (including $\eta > \lambda$), Lemma~\ref{lem:robust-frugality} establishes the robust bound.
\end{proof}

\section{Conclusion and Open Problems}
% In this work, we studied the design of frugal procurement auctions for the strategic uncapacitated facility location problem. We first established a tight frugality ratio of $3$ for the classic VCG auction. Prior to our work, the best known upper bound for VCG was $4$~\cite{talwar2003price} and there was no known lower bound. We then considered the problem in the learning-augmented framework where the auction is provided with predictions regarding the costs of the facilities. We designed a novel truthful auction that achieves a frugality ratio of $1+\epsilon$ when the predictions  are accurate, while maintaining a constant-factor robustness guarantee even when the predictions are arbitrarily wrong.
In this work, we studied the design of frugal procurement auctions for the strategic uncapacitated facility location problem. Our first main contribution establishes a tight frugality ratio of $3$ for the classic VCG auction. This result improves upon the best previously known upper bound of $4$~\cite{talwar2003price} and provides the first known lower bound for this problem. Complementing this classical analysis, we introduced a novel truthful mechanism in the learning-augmented framework. We showed that by leveraging predictions regarding facility costs, our auction achieves a frugality ratio of $1+\epsilon$ when predictions are accurate, while maintaining a constant-factor robustness guarantee against adversarial errors. While our analysis focuses on information-theoretic bounds, we further note that these mechanisms entail viable computational implementations via integer programs (see Appendix~\ref{app:computation}).

% Although the focus of our work is on the information-theoretic limitations of truthful auctions, we note that our proposed mechanisms  can be formulated as integer programs (IPs). Please see Appendix~\ref{app:computation} for the IP implementation.

Our work leaves several exciting directions open. In the classic setting, while we have settled the exact frugality ratio of the VCG auction, the fundamental limit of truthful auctions remains unknown. A key open question is whether VCG is optimal, or if an alternative design can achieve a frugality ratio strictly better than $3$. In the learning-augmented setting, an important next step is to fully characterize the Pareto-optimal frontier between consistency and robustness. Finally, there is significant promise in exploring more nuanced definitions of prediction error beyond simple magnitude—such as the number of mispredictions or the recent "mostly and approximately correct" (MAC) framework \cite{BGT24}. Analyzing mechanisms under such models could yield auctions that degrade more gracefully given imperfect, yet structurally informative, predictions.

\newpage
\acksection{
Eric Balkanski was supported by NSF grants CCF-2210501 and IIS-2147361. Nicholas DeFilippis was supported by an NSF Graduate
Research Fellowship under Grant No. DGE-2039655. Vasilis Gkatzelis and Xizhi Tan were supported by NSF CAREER award CCF-2047907 and NSF grant CCF-2210502. Any opinion, findings, and conclusions or recommendations expressed in this material are those of the authors and do not necessarily reflect the views of the National Science Foundation. 
}

\bibliographystyle{abbrvnat}
\bibliography{biblio}
\appendix
\section{Extended Related Work}\label{app:relatedwork}
\paragraph{Algorithms with Predictions.} 
In recent years, the learning‑augmented framework has become a leading paradigm for algorithm design and analysis. For an overview of its early developments, see \cite{MV22}, and for a comprehensive, up‑to‑date survey, consult \cite{alps}. By incorporating (potentially imperfect) predictions, this framework overcomes the limitations of classical worst‑case analyses. Indeed, over the past five years, hundreds of papers have revisited core algorithmic problems through this lens, with prominent examples including online paging \cite{lykouris2018competitive}, scheduling \cite{PSK18}, covering and knapsack‑constrained optimization \cite{BMS20, IKQP21}, Nash social welfare maximization \cite{banerjee2020online}, variations of the secretary problem \cite{AGKK23, DLLV21, KY23}, and a variety of graph‑based challenges \cite{azar2022online}.

\paragraph{UFL.} The metric Uncapacitated Facility Location (UFL) problem is a fundamental NP-hard optimization problem, widely studied due to its theoretical significance and practical relevance in operations research and algorithmic game theory. Early foundational results established the APX-hardness of the problem, showing that it cannot be approximated within a factor better than $1.463$ unless $P=NP$ \cite{guha1999greedy}. The first constant-factor approximation was provided by \citet{shmoys1997approximation}, who achieved a ratio of approximately $3.16$ . Subsequent improvements significantly narrowed the approximation gap, with notable progress by \citet{CS03} to 1.736 and \citet{mahdian2002improved} to $1.52$-approximation via sophisticated linear-programming techniques. \citet{byrka2010optimal} later improved this bound to a $1.50$-approximation using a novel bifactor algorithm . The current state-of-the-art algorithm, due to \citet{li2013approximating}, achieves an approximation ratio of $1.488$, leaving a small gap relative to the known hardness bound. 
% These milestones collectively underscore the theoretical depth of the UFL problem and highlight the subtle complexity involved in narrowing the approximation gap further.
% \section{Tightness of the Frugality Bound for VCG}\label{app:vcg}
% \input{appvcg}
% \section{Missing Proofs from Section~\ref{sec:UFL-with-advice}}\label{app:UFL-with-advice}
% \input{appLA}
% \section{Missing Proofs from Section~\ref{sec:error}}\label{app:error}
% \input{apperr}

\section{Computational Complexity}\label{app:computation}
 The standard IP formulation for UFL uses binary variables $x_\ell \in \{0, 1\}$ to denote whether facility $\ell \in L$ is opened and variables $y_{u\ell} \in [0, 1]$ for assigning user $u \in U$ to facility $\ell$.

The implementation of \texttt{Auction 1} proceeds as follows:
\begin{enumerate}
    \item \textbf{Determine the Predicted Optimal Set:} First, we solve a classic UFL integer program using the \textit{predicted} opening costs, $\{\hat{o}_\ell\}_{\ell \in L}$, to find the predicted optimal set, $\hat{OPT}$.

    \item \textbf{Find the Best Alternative Solution:} After identifying $\hat{OPT}$, its modified cost according to the auction rules is calculated directly. To find the best alternative, we solve a second UFL integer program using the \textit{true} opening costs, $\{o_\ell\}_{\ell \in L}$. A linear constraint is added to this IP to make the set $\hat{OPT}$ an infeasible solution:
    $$
    \sum_{\ell \in \hat{OPT}} x_\ell \le |\hat{OPT}| - 1
    $$
    This inequality ensures that at least one facility from the set $\hat{OPT}$ remains unopened, allowing the IP to find the optimal solution among all other feasible sets.

    \item \textbf{Select the Winning Set:} The auction's final output, $S^*$, is determined by comparing the modified cost of $\hat{OPT}$ with the cost of the best alternative solution found in the previous step. The set with the lower cost is chosen.

    \item \textbf{Compute Payments:} Finally, for each facility $\ell$ in the winning set $S^*$, its threshold payment is computed. This is done by solving one additional UFL integer program per winning facility to find the critical cost at which that facility would no longer be part of the optimal solution.
\end{enumerate}

\end{document}